\documentclass{jpp}
\usepackage{graphicx}
\usepackage[utf8]{inputenc}
\usepackage[T1]{fontenc}
\usepackage{algorithm}
\usepackage{algpseudocode}
\usepackage{amsmath}
\usepackage{dcolumn}
\usepackage{relsize}
\usepackage{bm}

\usepackage[utf8]{inputenc}
\usepackage[T1]{fontenc}
\usepackage{mathptmx}
\usepackage{etoolbox}
\usepackage{caption}
\usepackage{subcaption}
\usepackage{xcolor}

\usepackage{bm}

\usepackage[normalem]{ulem} 
 
 \definecolor{red}{rgb}{1.0,0.0,0.0}
 \definecolor{gre}{rgb}{0.5,0.5,1.5}
 \definecolor{blu}{rgb}{0.0,0.0,1.0}

\title{Fast particle trajectories and integrability in quasiaxisymmetric and quasihelical stellarators}

\author{Amelia Chambliss\aff{1}
  \corresp{\email{ac5114@columbia.edu}},
  E. Paul\aff{1}
 \and S. R. Hudson\aff{2}}

\affiliation{\aff{1}Department of Applied Physics and Applied Mathematics, Columbia University, New York NY 10025, USA
\aff{2}Theory Department, Princeton Plasma Physics Laboratory, Princeton NJ 08540, USA}

\begin{document}
\maketitle

\begin{abstract}
Even if the magnetic field in a stellarator is integrable, phase-space integrability for energetic particle guiding center trajectories is not guaranteed. Both trapped and passing particle trajectories can experience convective losses, caused by wide phase-space island formation, and diffusive losses, caused by phase-space island overlap.
By locating trajectories that are closed in the angle coordinate but not necessarily closed
in the radial coordinate, we can quantify the magnitude of the perturbation that results in island
formation. We characterize island width and island overlap in quasihelical (QH) and quasiaxisymmetric (QA) finite-$\beta$ equilibria for both trapped and passing energetic particles. For trapped particles in QH, low-shear toroidal precession frequency profiles near zero result in wide island formation. While QA transit frequencies do not cross through the zero resonance, we observe that island overlap is more likely since higher shear results in the crossing of more low-order resonances.

\end{abstract}

\section{\label{sec:level1}Introduction}
 In stellarator reactor design, confinement of fast particles is crucial for maintaining the temperatures necessary for fusion \citep{White_2015}. Rapid loss of $\alpha$ particles can be detrimental to device walls \citep{heatflux}. The low collisionality of energetic particles produces heightened sensitivity to resonances for fast $\alpha$'s compared to the thermal bulk particles. 
 Resonance occurs when the characteristic frequency of particles is rational, resulting in closed orbits \citep{Helander_2014,Rodríguez_Mackenbach_2023}. 
 Resonant perturbations result in the formation of phase-space islands for energetic particles. Particle energy and magnetic moment determine the frequency of orbits, and correspondingly the potential for drift island formation \citep{White_2022}, which can cause losses via radial convective transport and radial  transport.\\
 \indent Radial convective transport is caused by either large drift island formation resulting in resonant convection, or misalignment between drift surfaces and flux surfaces, known as drift surface convection \citep{Paul_2022,Diffusive}. For trapped particles, radial convective transport is defined by monotonic drift radially outward of the banana tips \citep{Paul_2022}. For passing particles, radial convective transport occurs when a passing trajectory experiences directed radial drift of the guiding center \citep{Paul_2022}. In the case of resonant convection, wide islands result in radial particle drifts due to island intersections with the boundary. Drift surface convection for trapped particles, also known as superbanana transport, has been characterized for stellarators by the $\Gamma_c$ metric in KNOSOS \citep{Velasco_2021}. In this case, misalignment of drift surfaces with flux surfaces can result in drift surfaces intersecting the boundary \citep{Nemov}. For trapped particles in tokamaks, guiding-center tracing shows rapid outward radial drifts via banana-harmonic resonances \citep{PhysRevLett.110.185004} which have been shown experimentally to inhibit toroidal rotation in KSTAR \citep{PhysRevLett.111.095002}, but the impact of convective transport in stellarators is insufficiently understood. A method compatible with non-axisymmetric geometries to characterize both types of radial convective transport for trapped and passing particles is necessary to better analyze radial convective losses in stellarators.\\
 \indent Another significant loss mechanism for energetic particles is diffusive radial transport, which results from chaotic regions in phase-space induced by island overlap. For trapped particles, diffusive radial transport is defined by chaotic motion of banana tips. For passing particles, chaotic orbits can lead to radial diffusion. Model map analysis has been used to analyze diffusive transport for trapped particles in tokamaks \citep{Goldston}. Diffusive transport for passing particles in TFTR has been visualized using Poincar\'e maps of guiding-center orbits, demonstrating stochasticity and island overlap \citep{mynick}. Phase-space Poincar\'e maps demonstrating chaos for passing particles in stellarators have been developed, but measures of island overlap and stochasticity are neglected \citep{White_2022}. As with convective transport, these existing methods used for diffusive transport analysis in tokamaks must be extended to understand the impact of this loss mechanism in general stellarator geometries.\\
 \indent Evaluation of distance from integrability can be useful for characterizing both convective and diffusive loss mechanisms. Integrability analysis of Hamiltonian systems in phase-space has been conducted for DIII-D subject to  Alfv\'enic perturbations \citep{White_2011}. Detection of non-integrability for surfaces of a given topology was enabled by rotation of a phase vector defined between adjacent orbits to evaluate the presence of phase-space islands and define stochastic regions. Integrability analysis was extended to quasiaxisymmetric (QA) stellarators using NEO-RT, an adapted tokamak code \citep{Albert_2022,NEORT}. The integrable Hamiltonian is defined for the underlying exact QA field, while the field perturbation from QA corresponds to the non-integrable Hamiltonian. The resonance of the perturbed system is then used to analytically compute island width using the thin-orbit assumption \citep{Albert_2022}. However, application of integrability analysis to other types of quasisymmetry with inclusion of wide orbits would be useful in guiding design.\\
 \indent While optimization for proxies of guiding-center confinement such as quasisymmetry have greatly improved stellarator performance \citep{QA,QH_vac}, symmetry metrics do not account for energetic particle phase-space island formation. Analytic work defines the bounce-averaged drift frequencies for trapped particles using the near-axis expansion in quasisymmetry to evaluate quasisymmetric field stability through analysis of maximum-$J$ \citep{Rodríguez_Mackenbach_2023}, which is related to energetic particle confinement. Equilibria have been optimized using metrics that incorporate some energetic particle physics, including $\Gamma_c$ for QA \citep{LeViness_2023,Bader_Drevlak_Anderson_Faber_Hegna_Likin_Schmitt_Talmadge_2019} and maximum-$J$ for a quasi-isodynamic case \citep{goodman2024quasiisodynamicstellaratorslowturbulence}, but drift islands and overlap are not accounted for directly. Investigation of resonances in general equilibria can inform future optimization for the avoidance of potentially harmful resonant perturbations.\\
 \indent There have been many different approaches in the literature for characterizing energetic particle transport mechanisms, but methods have not yet been applied to general stellarator geometries. Poincar\'e maps of guiding-center orbits have enabled analysis of both convective and diffusive transport for both particle classes \citep{Paul_2022}. Distance from integrability has been explored in tokamaks and in QA stellarators. However, in general stellarator equilibria, distance from integrability has yet to be evaluated for both convective and diffusive transport mechanisms for trapped and passing particles. Integrability analysis can be extended to general stellarator geometries in a way that is robust to island overlap, wide orbits, and regions of phase-space chaos.\\
\indent Since phase-space chaos can occur for both trapped and passing particles, one can describe distance from integrability for both systems. 
For guiding-center trajectories in a non-integrable system, which may exhibit phase-space chaos, it is useful to define a nearby integrable map. 
Motivated by the work of Dewar {\it et al.} on quadratic-flux minimizing surfaces \citep{DEWAR1994,HUDSON1998246,HudsonandDewar1999}, this can be achieved by locating pseudo-periodic curves.
Given a perturbed system, these orbits recover the periodic dynamics of the unperturbed system. 
The pseudo-periodic curve can then be used to describe the distance from integrability and evaluate drift island width and overlap.\\
\indent We define characteristic frequencies for trapped and passing trajectories to compute resonances and examine the impact of resonant perturbations on both QA and QH configurations. In Section \ref{sec:H_gc_motion}, we introduce the Hamiltonian for guiding-center motion to describe particle classes and the role of pitch angle in particle transport. We present the equilibria used in Section \ref{sec:equil}. We 
define area-preserving maps for the trapped and passing classes in Section \ref{sec:maps}, defining pseudo-periodic orbits to characterize distance from integrability. Computation of characteristic frequencies and an analytic simplification for trapped particle bounce frequency using the near-axis expansion are provided in Section \ref{sec:char_freqs}. Using our measure of distance from integrability, we produce analytic expressions for island widths in both cases and compare with the results of our mapping in Section \ref{sec:qfm}. A qualitative investigation of island overlap for a QA case far from integrability is also provided. In Section \ref{sec:results} we present our results. We conclude and discuss potential future investigations and open questions in Section \ref{sec:conclusion}.\\

\section{The guiding-center Hamiltonian}
\label{sec:H_gc_motion}
\begin{figure}
    \centering
    \includegraphics[width=0.6\textwidth]{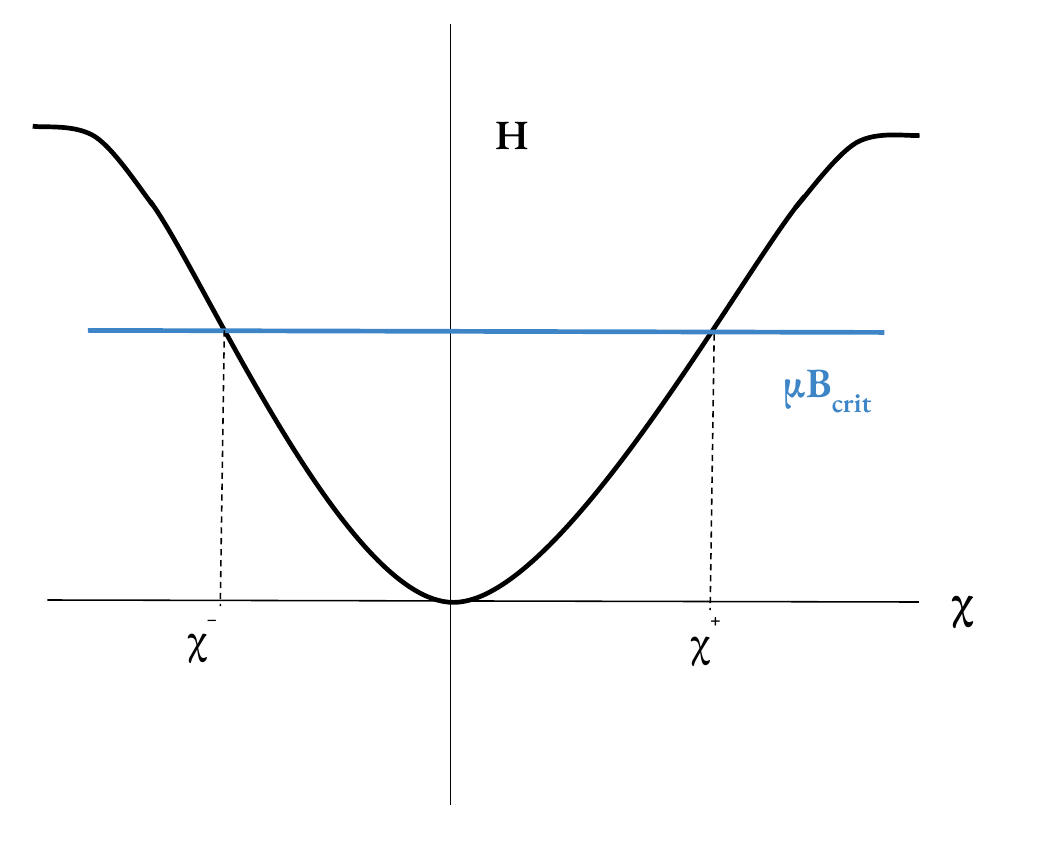}
    \caption{The effective potential at constant $s$ determined by $\mu B$. $\chi_{-}$ and $\chi_+$ are the mirror points along the $\chi$ axis, and $\mu B_{\text{crit}}$ the total energy for the particle.}
    \label{fig:potl}
\end{figure}

The Hamiltonian for guiding-center motion is given by \citep{Littlejohn_1983}:
\begin{equation}
    H(s,\theta,\zeta,v_{\|})=\textstyle\frac{1}{2} mv_{\|}^2 + \mu B,
\end{equation}
where $B$ is the magnetic field magnitude, $v_{\|}$ is the velocity of the particle parallel to the field, and $\mu$ is the magnetic moment. For particle motion, $H$ is conserved and equal to total particle energy.
This statement shows that $\mu B$ acts as an effective potential for a particle in a magnetic field. An illustration of this effective potential energy attribute is shown in Figure \ref{fig:potl}. For quasisymmetric configurations, field strength is a function of the helical angle $\chi=\theta-N\zeta$ for poloidal Boozer coordinate $\theta$, toroidal Boozer angle $\zeta$, and helicity $N$, which takes on value $0$ for QA configurations and $\pm N_{fp}$ for QH configurations where $N_{fp}$ is the number of field periods.
The helical angle $\chi$ is defined such that for quasisymmetric configurations, $B(s,\chi$) where $s$ is the normalized flux. As a particle encounters varying $B$ throughout its orbit, $B_{\text{crit}}$ is the field strength value at which a particle will bounce. The value of pitch angle $\lambda=\mu B_0/H$, where $B_0$ is the normalizing field strength on-axis, will determine the effective potential and the value of $B_{\text{crit}}$.
\section{Equilibria}
\label{sec:equil}
\begin{figure*}
    \centering
    \includegraphics[width=0.6\textwidth]{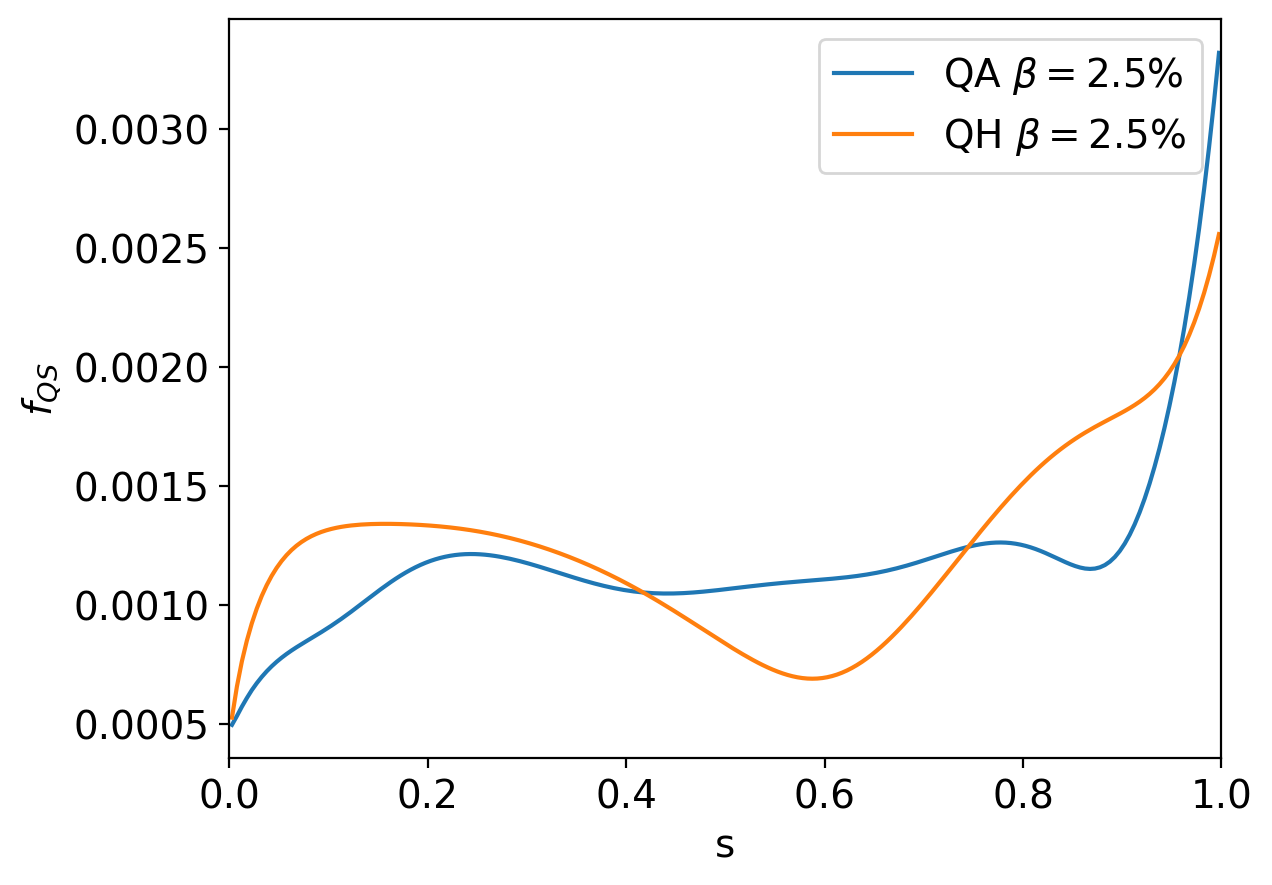}
    \caption{Quasisymmetry error $f_{QS}$ for configurations used. For the two configurations with $\beta=2.5\%$, we see similar magnitudes of $f_{QS}$, but very different transport and resonance sensitivities will be demonstrated.}
    \label{fig:QS}
\end{figure*}

We perform our analysis on reactor-scale configurations with finite-$\beta$ for $\beta=\frac{2\mu_0\langle p \rangle}{B^2}$ where $\langle p \rangle$ is the mean plasma pressure and $\mu_0$ is the vacuum magnetic permeability constant. We use QA and QH configurations with $\beta=2.5\%$ \citep{QA_config}. These equilibria were produced with VMEC, and therefore have no magnetic islands. Distance from quasisymmetry for these configurations can be evaluated by summing the symmetry-breaking Fourier modes \citep{QH_vac}:
\begin{equation}
    f_{QS}=\mathlarger{\mathlarger{\sum}}_{m, n\neq Nm/N_{fp}}\left(\frac{B_{m,n}}{B_{0,0}}\right)^2
\end{equation}
for integers $n, m$. Field strength is defined such that
\begin{equation}
    B(s,\theta,\zeta)=\mathlarger{\mathlarger{\sum}}_{m,n}B_{m,n}(s)\cos{(m\theta-nN_{fp}\zeta)}.
\end{equation}
The quasisymmetry metrics for these configurations are shown in Figure \ref{fig:QS}. Despite both finite-$\beta$ equilibria possessing similar $f_{QS}$ magnitudes, we demonstrate in later sections that resonances of fast particle drift frequencies and resulting perturbation-induced transport differ significantly for these configurations.

\section{Map computations}
\label{sec:maps}
To visualize drift islands and guiding-center transport in these equilibria, we can construct area-preserving phase-space maps for trapped and passing particles.
For a chosen pitch, each guiding-center trajectory has four dimensions: parallel velocity $v_{\|}$, as well as three spatial dimensions in ($s,\chi,\zeta$).
For each particle class, this four-dimensional problem must be reduced to two dimensions for visualization. Dimension reduction depends on particle class. Energy $H$ and pitch $\lambda$ are fixed for each trajectory. 
A map can be produced for each pitch value. Particles are lost when their trajectory leaves the boundary, and trajectories that travel too close to the magnetic axis are eliminated to avoid a coordinate singularity.
\subsection{Passing particles}
For passing particles, we define particle initial points in phase-space at a chosen ($s_0,\chi_0,v_{\|}$) at $\zeta_0=0$. We define the mapping
\begin{equation}
    \mathcal{M}_p(s_0,\chi_0)\longrightarrow (s',\chi ')
    \label{eq:p_map}
\end{equation}
that advances a particle trajectory from its initial location to the resulting $(s',\chi')$ point in the $\zeta=0$ plane. Through asserting that $v_{\|}$ does not change sign, we eliminate bouncing particles and use $H$-conservation and $\lambda$ to determine $v_{\|}$ given $(s,\chi,\zeta)$, enabling the two-dimensional expression of (\ref{eq:p_map}). 
This $v_{\|}$ condition ensures that we capture each trajectory in unidirectional motion only, analogous to field-line flow Poincar\'e plots.

\subsection{Trapped particles}
For trapped particles, $\lambda$ is chosen and particles are initialized at the bounce point where $B=B_{\text{crit}}=B_0/\lambda$. Under the assumption that particles remain trapped in a single well, the mirror point in $\chi$ is determined given $s$ and $\zeta$ by seeking the value of $\chi$ where $B=B_{\text{crit}}$. The mapping is thus defined
\begin{equation}
    \mathcal{M}_t(s_0,\zeta_0) \longrightarrow (s',\zeta').
    \label{eq:t_map}
\end{equation}
We apply (\ref{eq:t_map}) by following guiding-center trajectories until $v_{\|}$ has passed through zero for both $v_{\|}'>0$ and $v_{\|}'<0$, isolating a full bounce period.
The second $v_{\|}=0$ crossing corresponds to the ($s,\chi,\zeta$) point along a trajectory where $B=B_{\text{crit}}$, defined as $(s',\chi',\zeta')$. The single-well assumption is used to further reduce dimensionality. This eliminates consideration of ripple-trapped and transitioning particles. These mirror points are then plotted in the $(s,\zeta)$-plane.

\section{Characteristic frequencies}
\label{sec:char_freqs}
To determine the resonance conditions for both particle classes, we must first define the frequencies associated with particle motion for both trapped and passing particles. For passing particles, effective helicity is given by $\omega_{\chi}$, the displacement in $\chi$ per displacement in $\zeta$ averaged over many toroidal transits. For the mapping in (\ref{eq:p_map}),
\begin{equation}
    \omega_{\chi} = \frac{1}{2\pi} \left\langle \oint \dot\chi dt \right\rangle,
    \label{eq:wchi}
\end{equation}
since passing trajectories travel a full rotation of $2\pi$ in $\zeta$ after one map application. The average over many transits is indicated by $\langle \rangle$ and $\oint$ denotes integration over one toroidal transit.
For the mapping in (\ref{eq:t_map}), trapped trajectories have toroidal precession frequency $\omega_{\zeta}$. This is given by the normalized change in toroidal angle $\zeta$ after a full bounce period, defined as
\begin{equation}
    \omega_{\zeta} = \frac{1}{2\pi} \left\langle \oint \dot\zeta dt \right\rangle,
    \label{eq:wzeta}
\end{equation}
for integration over the full bounce period in $\zeta$. The average over many bounce periods is denoted by $\langle \rangle$.
\subsection{Resonances}
\begin{figure*}
    \centering
    \begin{subfigure}{0.4\textwidth}
        \centering
        \includegraphics[width=\textwidth]{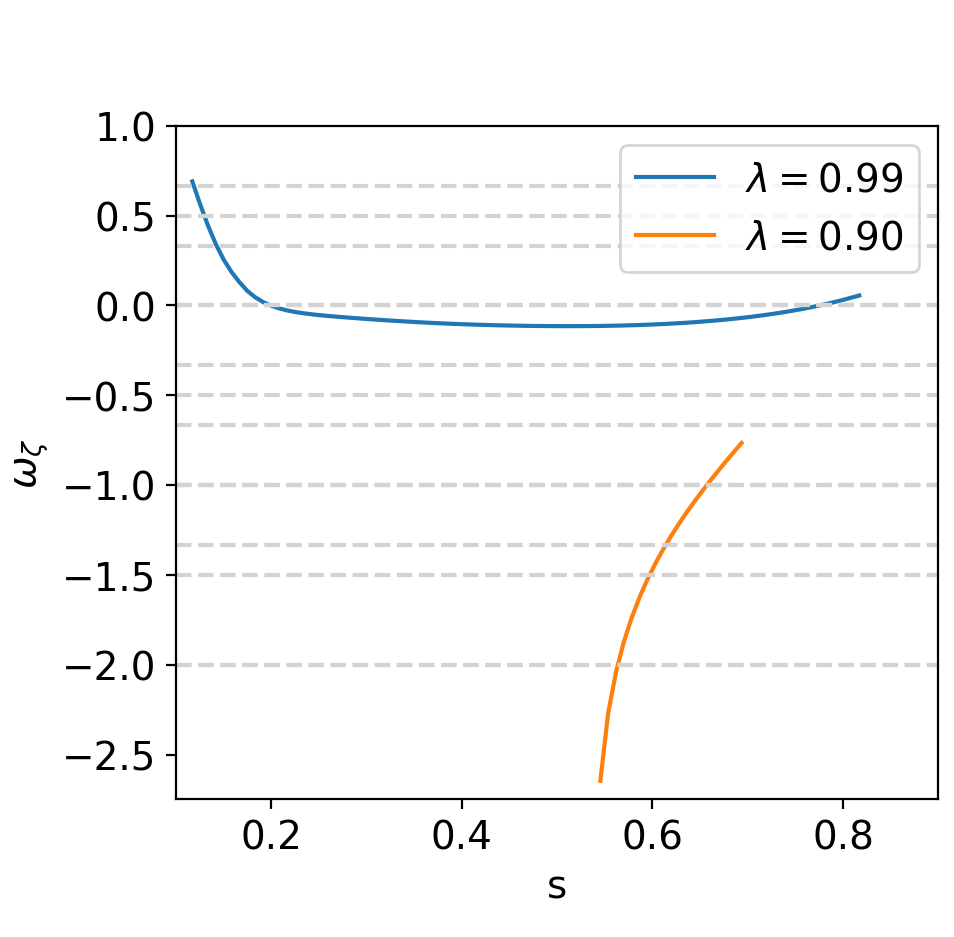}
        \caption{$\omega_{\zeta}$ for trapped particles in QA with different values of $\lambda$ spanning the pitch range. Low-order resonances up to $q=3$ are plotted.}
        \label{fig:freq_qab_t}
    \end{subfigure}
    \hfill
    \begin{subfigure}{0.4\textwidth}
        \centering
        \includegraphics[width=\textwidth]{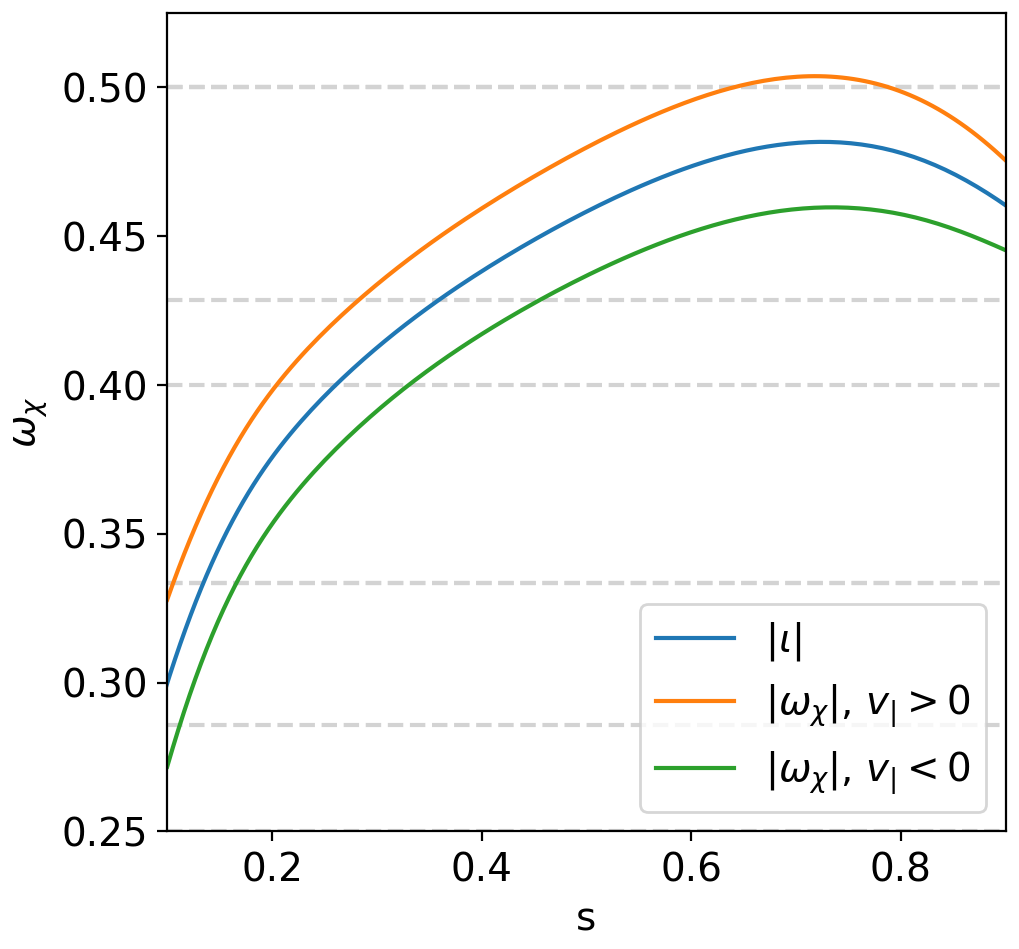}
        \caption{$\omega_{\chi}$ for co-propagating and counter-propagating passing particles in QA. Resonances up to $q=7$ are plotted.}
        \label{fig:freq_qab_p}
    \end{subfigure}
    \begin{subfigure}{0.4\textwidth}
        \centering
        \includegraphics[width=\textwidth]{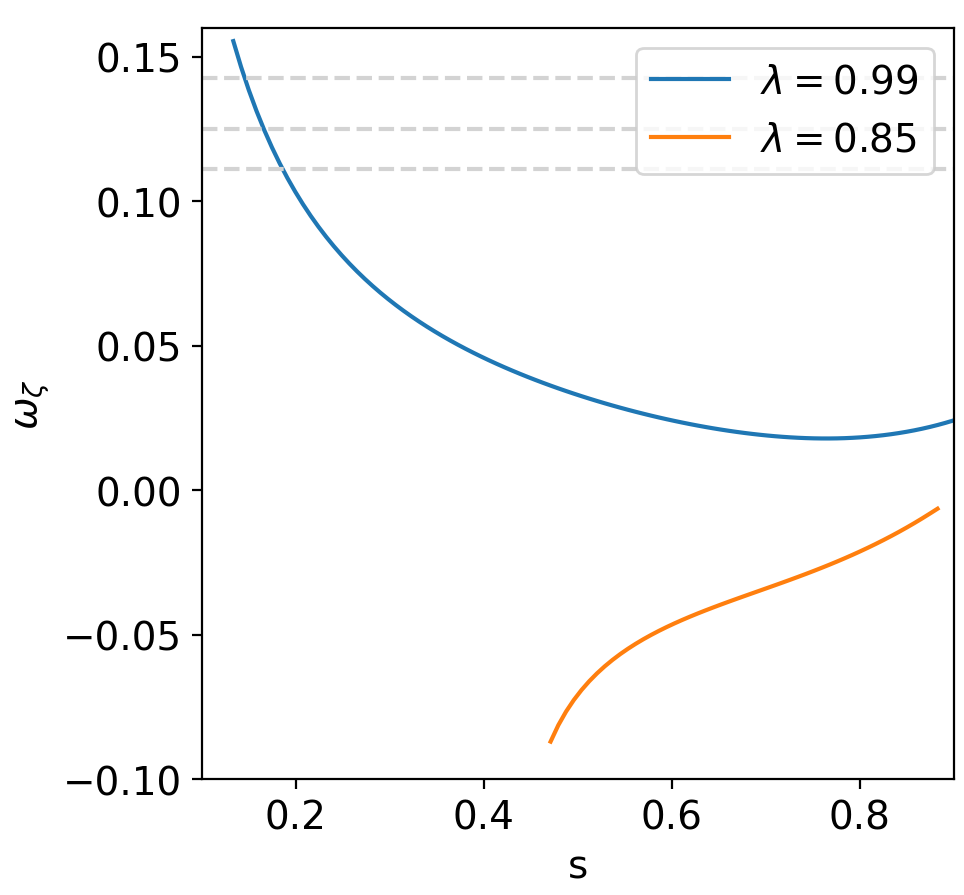}
        \caption{$\omega_{\zeta}$ for trapped particles in QH with different values of $\lambda$ spanning the pitch range. Low order resonances up to $q=9$ are plotted.}
        \label{fig:tfreq_qhb_t}
    \end{subfigure}
    \hfill
    \begin{subfigure}{0.4\textwidth}
        \centering
        \includegraphics[width=\textwidth]{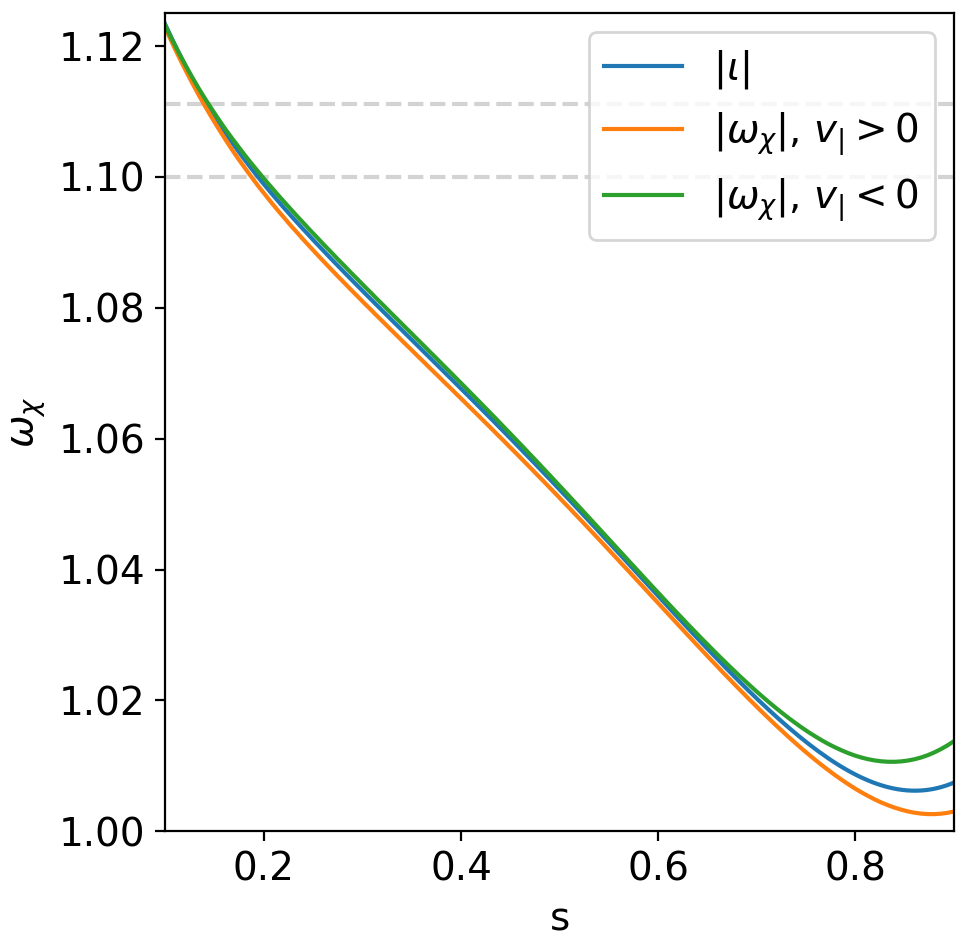}
        \caption{$\omega_{\chi}$ for co-propagating and counter-propagating passing particles in QH. Resonances up to $q=10$ are plotted.}
        \label{fig:freq_qhb_p}
    \end{subfigure}    \caption{Frequency profiles for trapped and passing trajectories for representative values of pitch in QH $\beta=2.5\%$ and QA $\beta=2.5\%$ equilibria. The range of frequency profiles is determined by upper and lower bounds of the pitch range for each equilibrium. Low-order resonances are plotted as dashed grey lines for each case. For passing particles, comparison to the $\iota$ profile is provided.}
    \label{fig:all_frequencies}
\end{figure*}

\begin{figure*}
    \centering
    \begin{subfigure}{0.45\textwidth}
        \includegraphics[width=\textwidth]{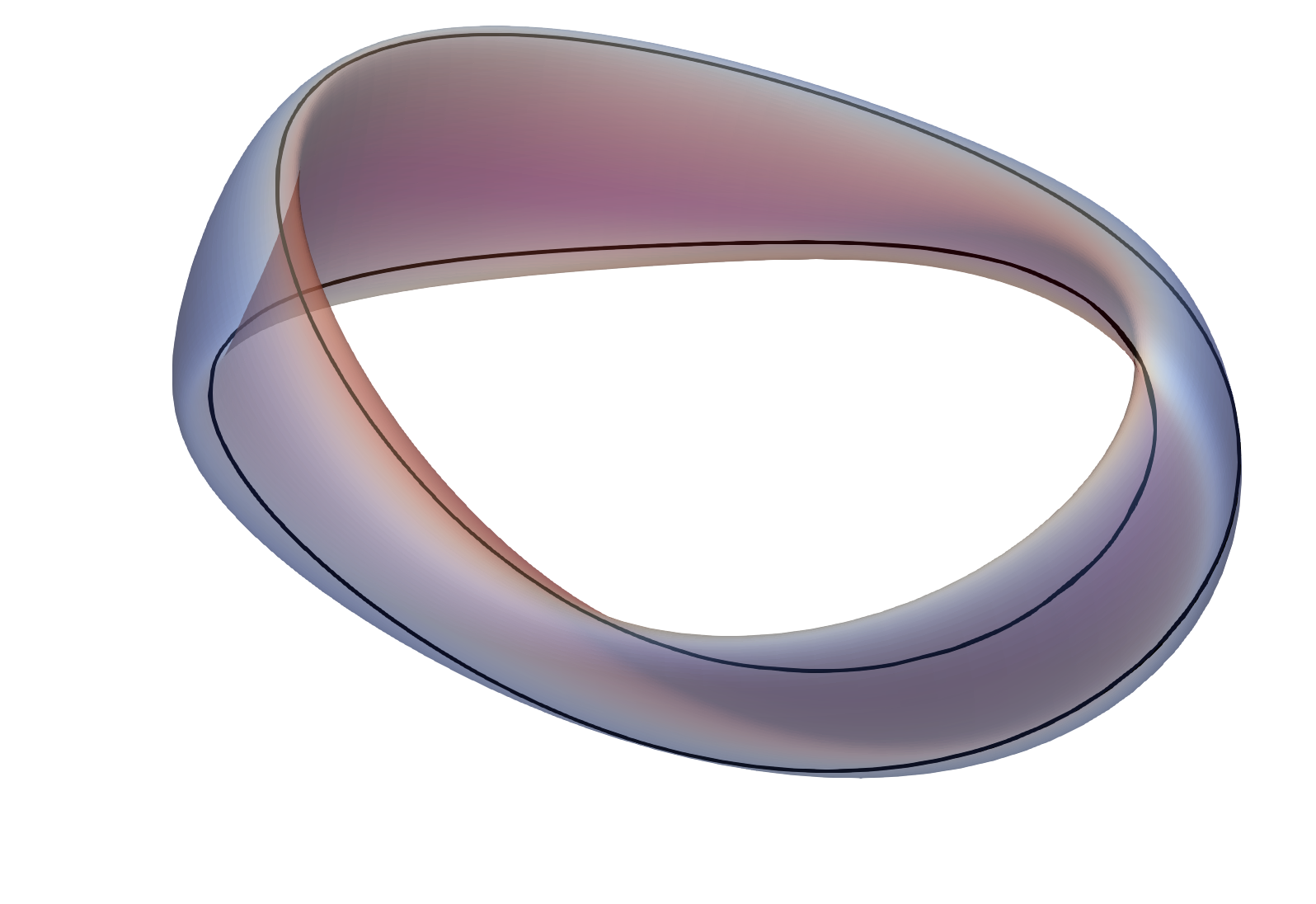}
        \caption{QA near $p=2$, $q=1$ resonance.}
        \label{fig:traj-qa_p}
    \end{subfigure}
    \hfill
    \begin{subfigure}{0.45\textwidth}
        \includegraphics[width=\textwidth]{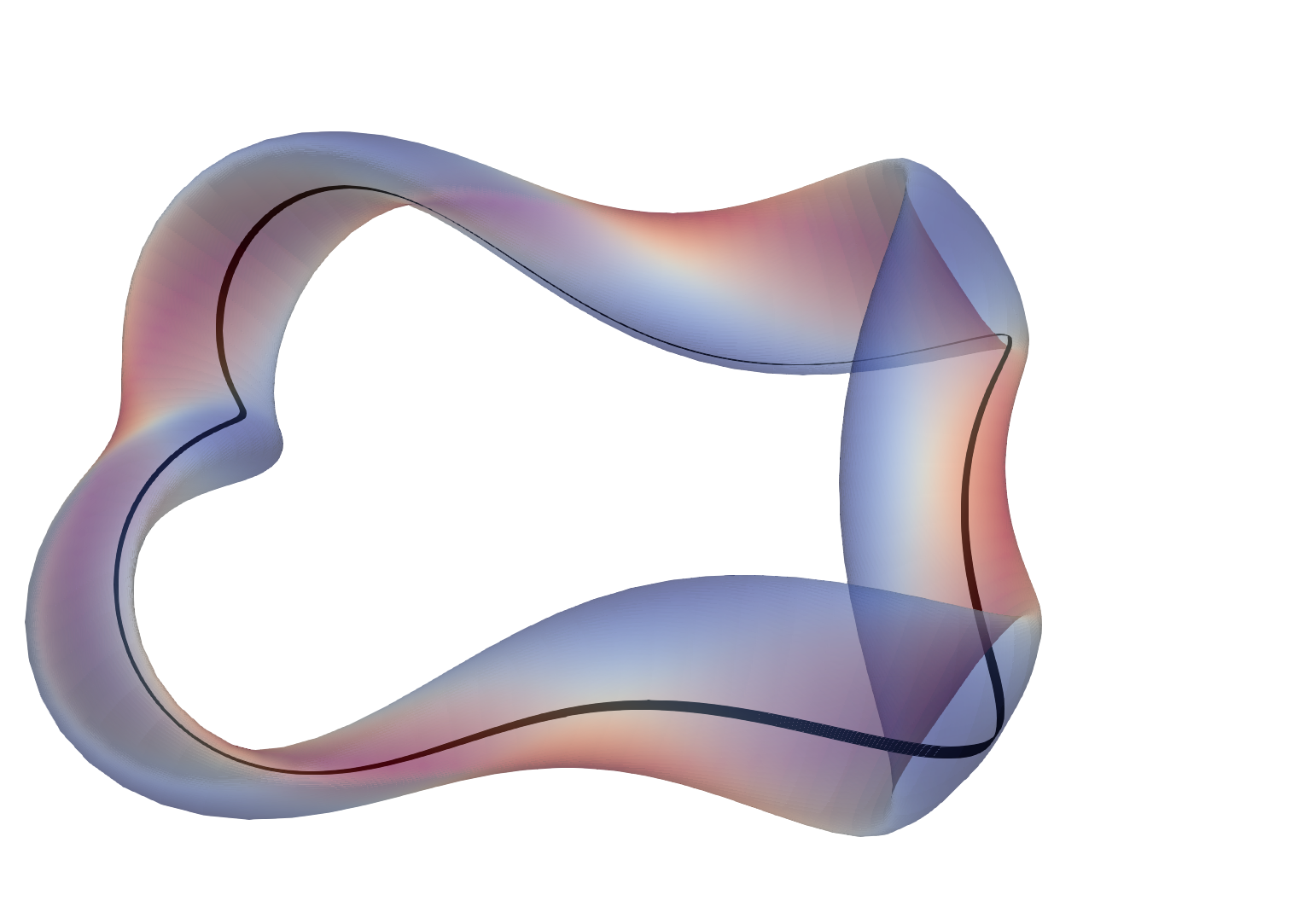}
        \caption{QH near the $p=-4$, $q=4$ resonance.}
        \label{fig:traj-qh_p}
    \end{subfigure}
    \caption{Passing particle trajectories close to resonance in QA and QH stellarators. Field strength on the last closed flux surface is indicated in color.}
    \label{fig:traj-passing}
\end{figure*}

\begin{figure*}
    \centering
    \begin{subfigure}{0.45\textwidth}   \includegraphics[width=\textwidth]{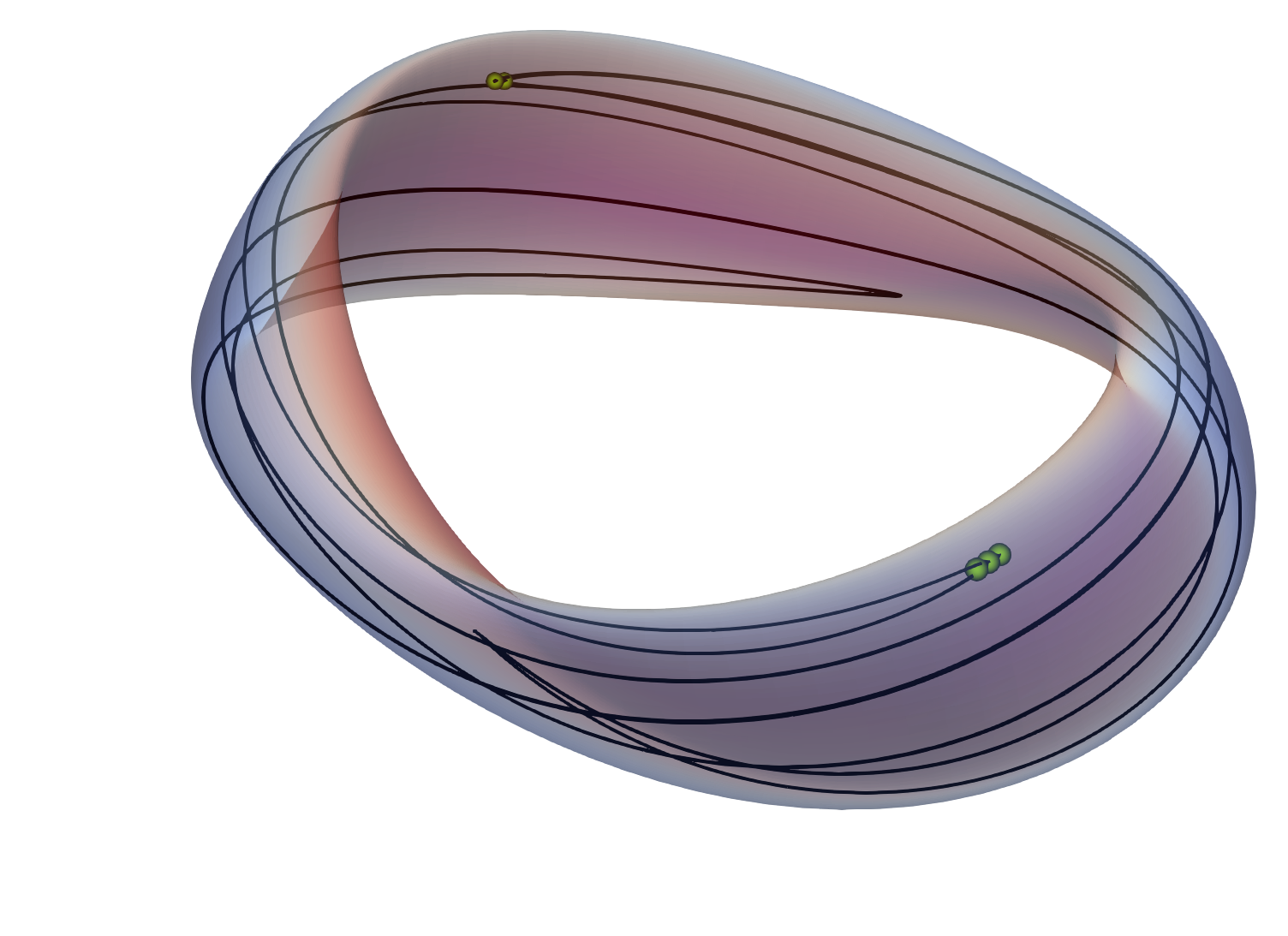}
        \caption{QA near $p=-1$, $q=2$ resonance.}
        \label{fig:traj-qa_t}
    \end{subfigure}
    \hfill
    \begin{subfigure}{0.45\textwidth}
        \includegraphics[width=\textwidth]{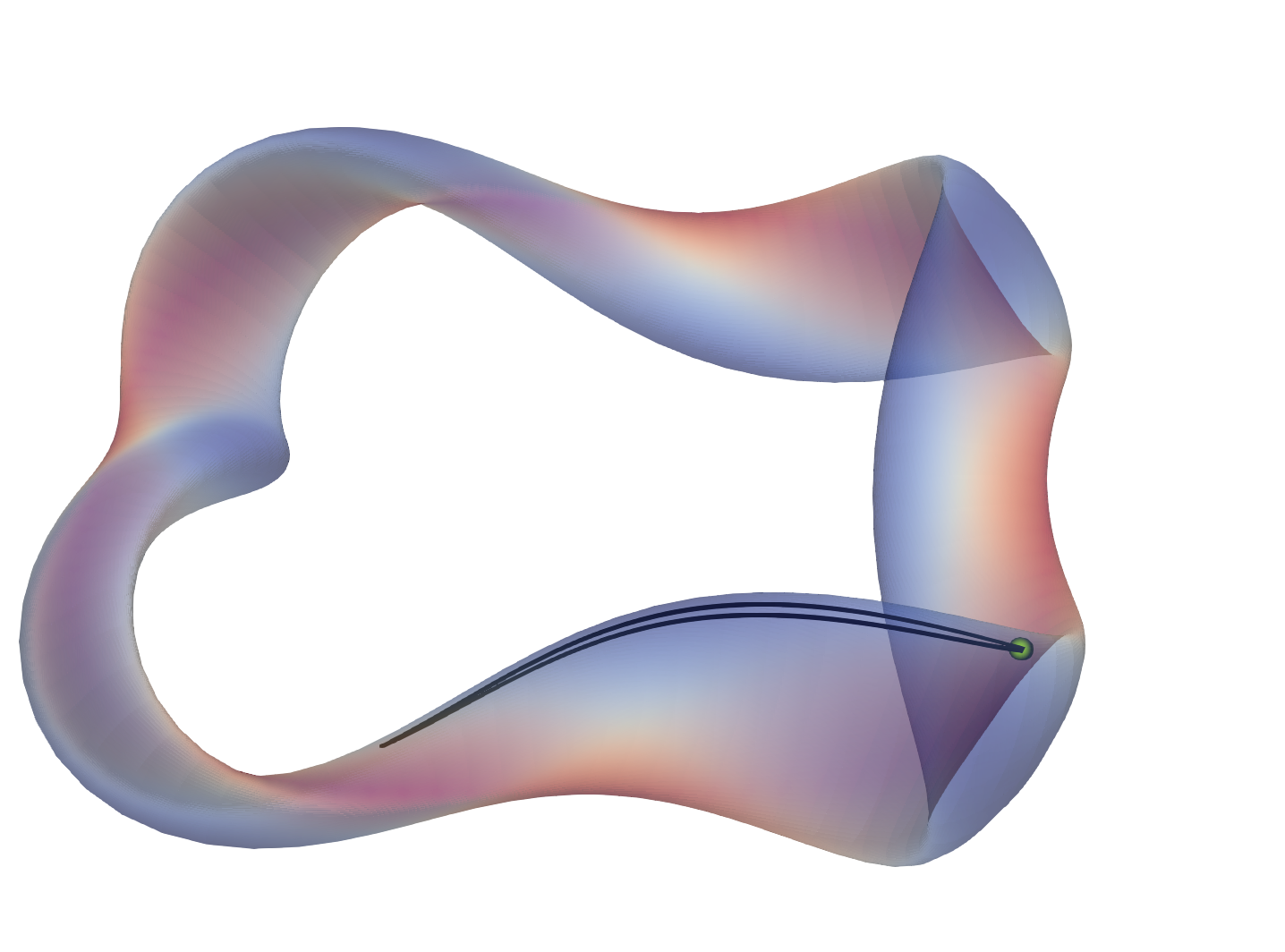}
        \caption{QH near the $p=0$, $q=1$ resonance.}
        \label{fig:traj-qh_t}
    \end{subfigure}
    \caption{Trapped particle trajectories close to resonance in QA and QH stellarators. Field strength on the last closed flux surface is indicated in color. Bounce points are shown in green.}
    \label{fig:traj-trapped}
\end{figure*}

For these characteristic frequencies, resonance conditions must be considered to evaluate relevant transport mechanisms associated with potential resonant perturbations. The resonance condition for trapped or passing particle orbits is
\begin{equation}
    \omega_{\phi}=p/q
    \label{eq:resonance}
\end{equation} for $\omega_{\phi}=\omega_{\chi}$ or $\omega_{\zeta}$. In other words, a particle orbit close to a resonance will nearly close on itself after $q$ map applications. 
When the frequency of a particle orbit meets (\ref{eq:resonance}), the orbit will be highly susceptible even to small perturbations in quasisymmetry,
and is more likely to be subject to convective or diffusive losses, particularly for low-order resonances.\\
\indent Numerical results for $\omega_{\zeta}$ and $\omega_{\chi}$ are shown for trapped and passing particles in QA and QH equilibria at representative values of pitch in Figure \ref{fig:all_frequencies}, with low-order resonance  crossings. A guiding-center tracing routine in SIMSOPT \citep{SIMSOPT} is used for fusion-born $\alpha$ particles in the reactor-scale equilibria shown in Figure \ref{fig:QS} \citep{Paul_2022}. We initialize particles at equidistant $s$ values at a constant angle $\phi=\chi$ or $\zeta$ depending on particle class. Frequencies are averaged over the guiding-center motion and symmetry-breaking modes are filtered out to isolate the frequencies of the unperturbed motion. For a perfect quasisymmetric case, the canonical momentum $P_{\zeta}(s, \chi)$ is conserved. Since trajectories are initialized at $\chi=0$, characteristic frequencies can be expressed as functions of $s$. For trapped trajectories, profiles were produced for pitch values close to the trapped-passing boundary and the deeply-trapped limit to demonstrate the range of frequency profiles present for the equilibrium. For passing trajectories, co-propagating and counter-propagating frequency profiles are shown with comparison to the rotational transform $\iota$. At low energies in QA, we expect agreement between $\omega_{\chi}$ and $\iota$, but variation is observed since drifts enable deviation of passing trajectories from field-line flow. As a result, when optimizing for a desirable $\iota$, passing particle flow resonances can differ from those of the field structure. Resonant orbits in 3D are shown in Figures \ref{fig:traj-passing} and \ref{fig:traj-trapped}. For passing particles, the orbit visibly closes on itself, while for trapped particles, bounce points return to the nearly the same location after $q$ bounce periods.

\subsection{Near-axis approximation}

\begin{figure}
    \centering
    \includegraphics[width=0.5\textwidth]{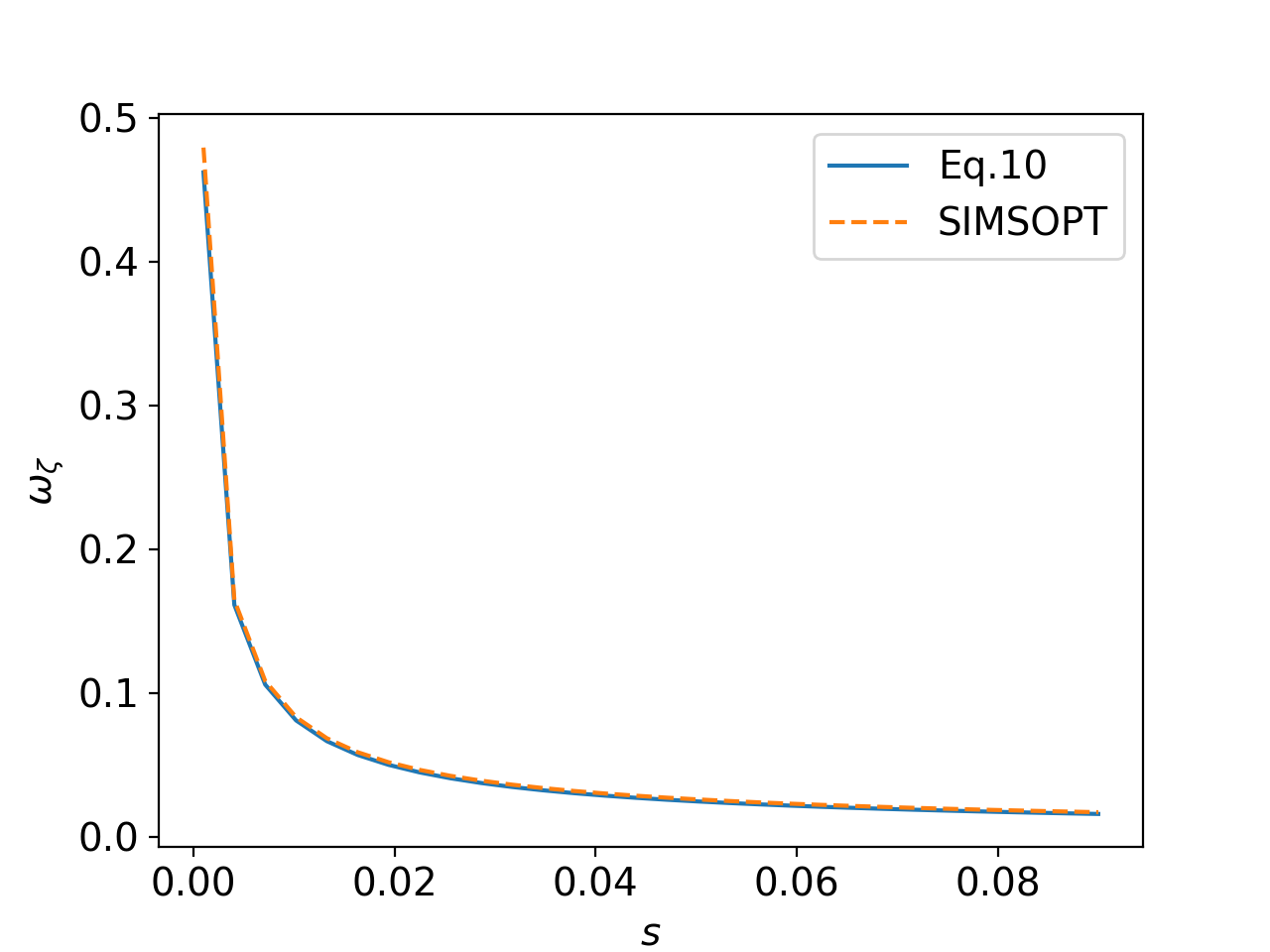}
    \caption{Comparison between the analytic form for $\omega_{\zeta}$ given by (\ref{eq:omega_zeta}) and numerical results from guiding-center tracing for $\omega_{\zeta}$ for a near-axis field.}
    \label{fig:freqmatch}
\end{figure}

\begin{figure}
    \centering
    \includegraphics[width=0.5\textwidth]{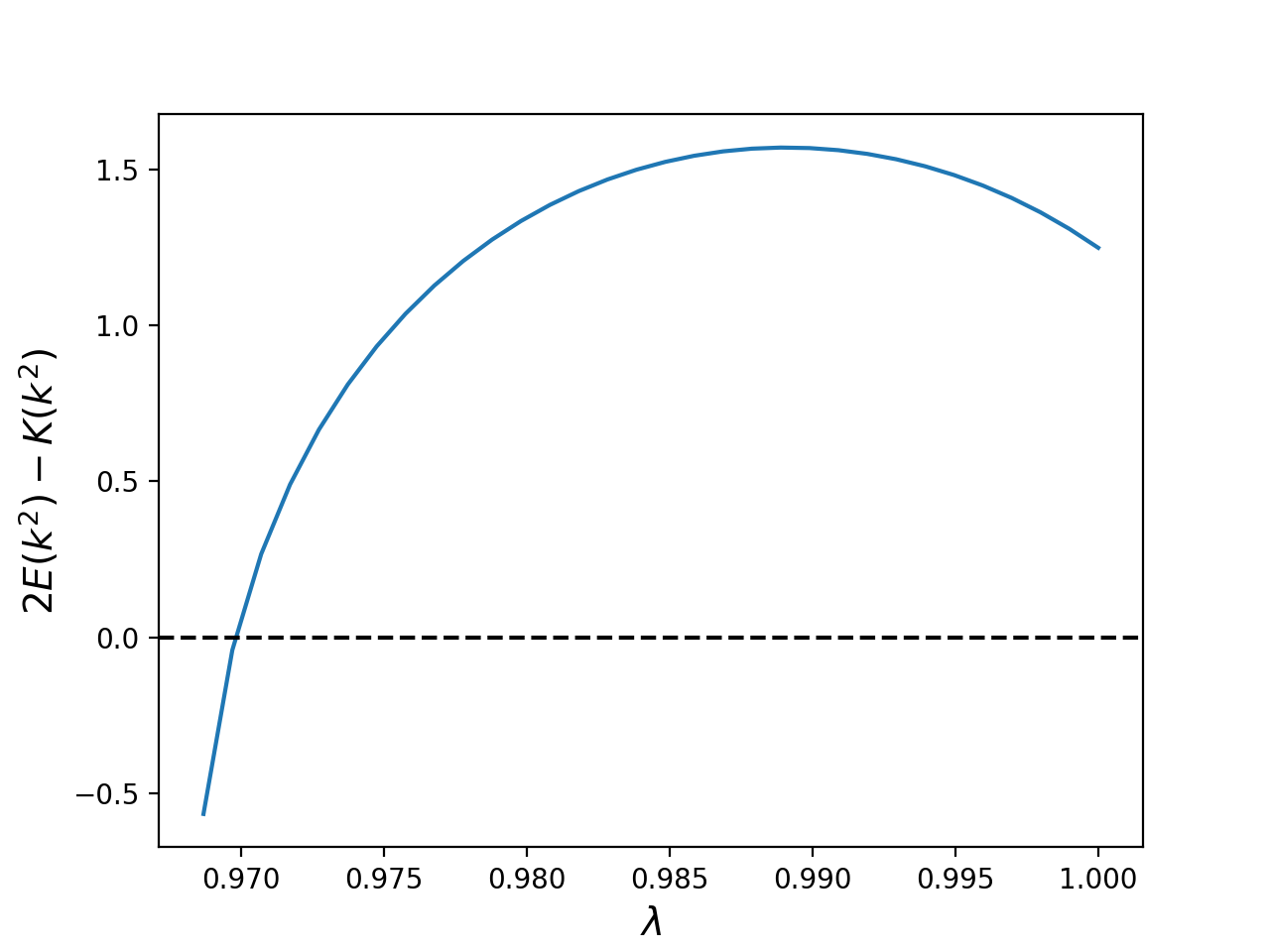}
    \caption{The value of (\ref{eq:omega_zeta}) versus pitch. The resonance at zero is indicated by the dashed line.}
    \label{fig:elliptic_integrals}
\end{figure}

\begin{figure*}
    \centering
    \begin{subfigure}{0.47\textwidth}
        \includegraphics[width=\textwidth]{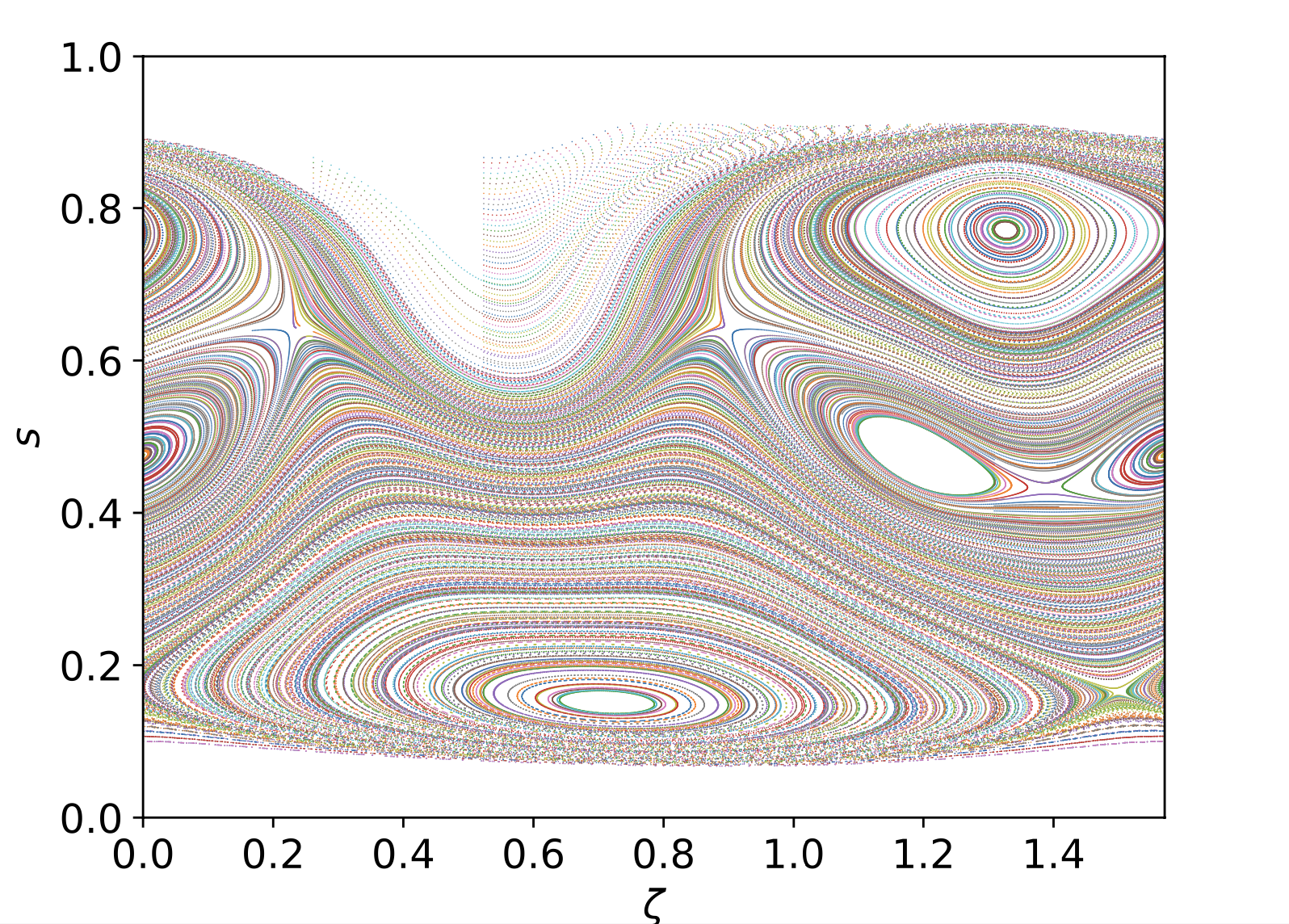}
        \caption{Poincar\'e map.}
        \label{fig:qhb_tr_map}
    \end{subfigure}
    \begin{subfigure}{0.47\textwidth}
        \includegraphics[width=\textwidth]{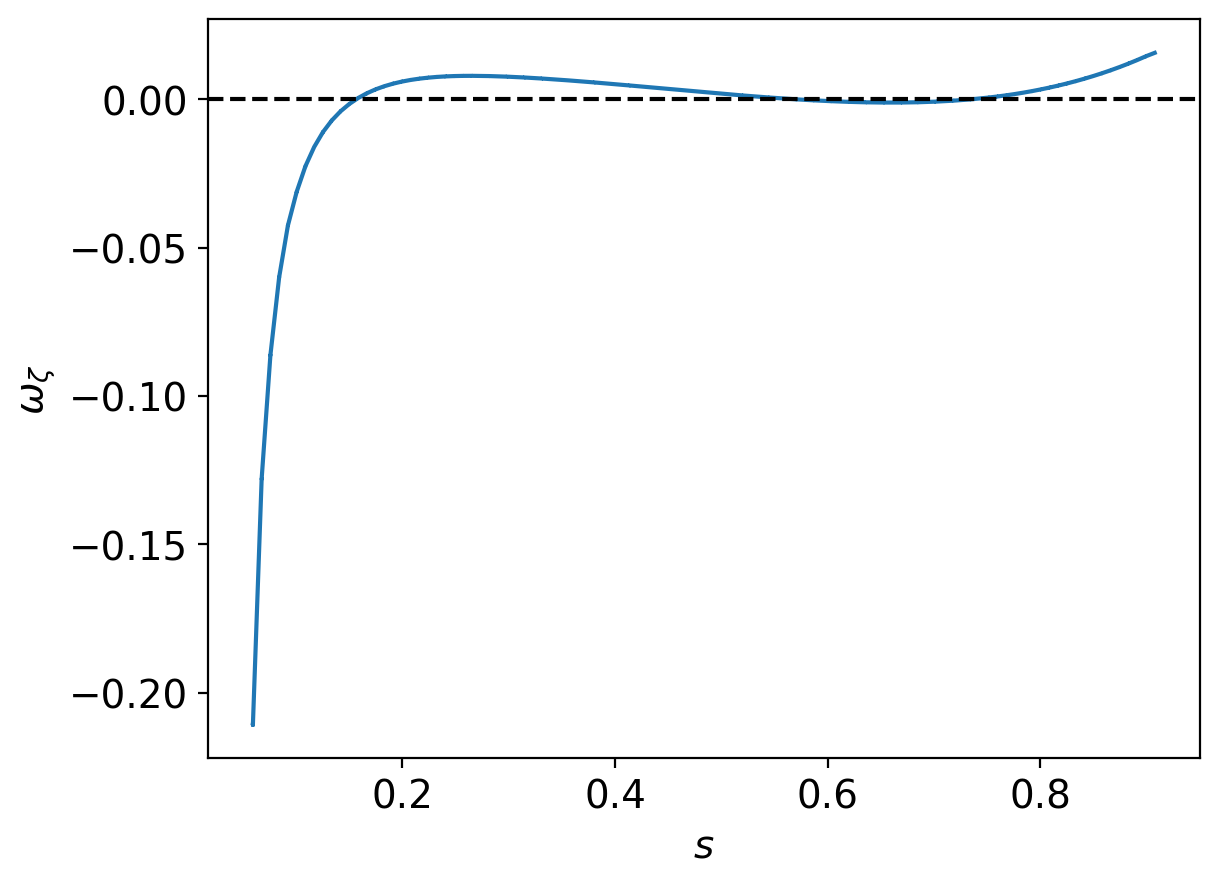}
        \caption{Frequency profile.}
        \label{fig:qhb_tr_freq}
    \end{subfigure}
    \caption{Poincar\'e map and frequency profile for trapped particles in QH equilibrium with $2.5\% \beta$ at $\lambda=0.95$. Multiple zero-crossings in the frequency profile correspond to island structures that appear distinctly at three different radii, resulting in a definitively nontwist map structure \citep{DELCASTILLONEGRETE19961}}.
    \label{fig:qhb_trapped}
\end{figure*}

These frequency expressions can be simplified using the near-axis approximation for the field structure. In near-axis theory, the magnetic field strength to first order in the distance from the magnetic axis $r = \sqrt{2\psi/B_0}$ for flux $\psi$ is given by \citep{Landreman_Sengupta_2018}:
\begin{equation}
    B = B_0 \left[1 - r \bar\eta \cos(\chi)\right],
    \label{eq:bfield}
\end{equation}
where $\bar\eta$ is a constant. 
To lowest order, rotational transform is constant, $\iota = \iota_0$. For a configuration with helicity $N$, the toroidal transit frequency after one full bounce period \citep{Rodríguez_Mackenbach_2023} is then proportional to
\begin{equation}
        \omega_{\zeta}\propto \frac{1}{(\iota_0 -N)^2}\left[2E(k^2)-K(k^2)\right]
    \label{eq:omega_zeta}
\end{equation}
where
\begin{equation}
    k^2=\frac{1-\lambda+\lambda r \bar\eta}{2\lambda r \bar\eta}.
    \label{eq:k}
\end{equation}
More detail is provided in Appendix \ref{app:dzeta}. 
 In Figure \ref{fig:freqmatch}, (\ref{eq:omega_zeta}) is compared with the frequency profile for particles in a near-axis field geometry given in (\ref{eq:bfield}), indicating good agreement. Figure \ref{fig:elliptic_integrals} demonstrates the pitch for which $2E(k^2)-K(k^2)$ crosses through zero. Heightened sensitivity to the $\omega_{\zeta}=0$ crossing for trapped particles is suggested. This implies that additional care should be taken to reduce symmetry-breaking modes for trajectories that resonate with the zero-crossing. The $1/(\iota_0-N)^2$ scaling results in a larger magnitude of $\omega_{\zeta}$ for QA than for QH configurations, suggesting a higher shear for QA and a heightened sensitivity to the zero-crossing for QH due to comparatively lower shear.\\
\indent We observe the impact of the $\omega_{\zeta}=0$ crossing on particles in a QH equilibrium in Figure \ref{fig:qhb_trapped}. Three zero-crossings result in the formation of large islands, which can lead to substantial convective radial transport. This suggests that quasisymmetry error should be reduced for pitches such that the toroidal precession frequency exhibits resonance with the zero-crossing.\\

\subsection{Pseudo-periodic trajectories}
\label{sec:qfm}
\begin{figure}
    \centering
    \includegraphics[width=0.5\linewidth]{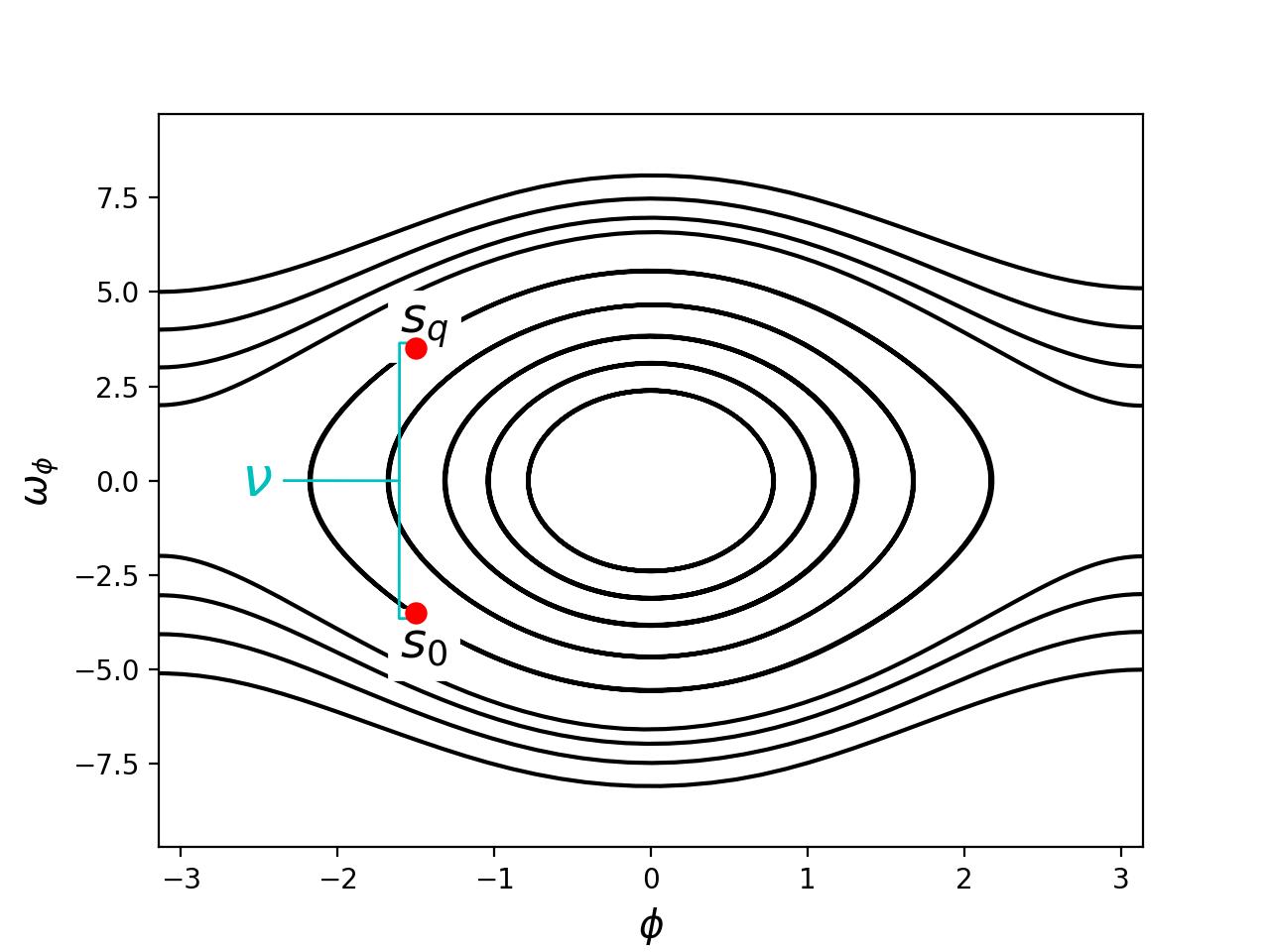}
    \caption{Schematic of the pseudo-periodic curve rootsolve.}
    \label{fig:psmap_schem}
\end{figure}

\indent Once perturbations are introduced, particle orbits are likely to become non-integrable, in that their orbits no longer lie on invariant surfaces in phase space.
For the non-integrable Hamiltonian system, it is useful to construct a nearby integrable Hamiltonian system, which can be used to decompose the given non-integrable Hamiltonian as an integrable Hamiltonian plus a small perturbation. 
Building on work by Dewar and co-workers \citep{DEWAR1994,Dewar_1995,DEWAR20122062,DEWAR199566}, we can construct a mapping that allows us to construct periodic orbits of the nearby integrable system. 
These orbits will pass through the X- and O-points of resonant perturbations, since they are stationary points of the action for this Hamiltonian system \citep{DEWAR199566}. 
Beginning with an initial guess $s_q$ for the radial location of the $p/q$ resonance at some angle $\phi_0$, the map is applied $q$ times. 
During these map applications, a root-solve is performed to recover a pseudo-periodic orbit closes in $\phi$ but not in $s$, arriving at $(s_0,\phi_0)$. 
The radial difference $\nu = s_q - s_0$ is a measure of the resonant perturbation that results in the island formation.
This is performed for uniformly spaced values of $\phi$. The pseudocode in Algorithm 1 describes this process, also illustrated in Figure \ref{fig:psmap_schem}. Radial distance $\nu (\phi)$ is a constant along each trajectory, and varies sinusoidally across the angle coordinate. This radial difference is related to the distance from integrability and can be used to evaluate island width.\\

\begin{algorithm}
\caption{Pseudocode for the discovery of pseudo-periodic curves}
\begin{algorithmic}[1]
\For{$p$, $q$}
    \State $s_q \gets \texttt{radialguess}(p/q)$
    \For{$\phi_i$ in $\left[\phi_0, \phi_n\right]$}
        \State Find $s_0$ such that:
        \State $\quad \mathcal{M}^q_{\nu}(s_0, \phi_i) = (s_q, \phi_i)$
        \State $\nu_i = s_q - s_0$
    \EndFor
\EndFor
\end{algorithmic}
\end{algorithm}

\indent For the passing map, our characteristic frequency is $\omega_{\chi}$. For a resonance $\omega_{\chi}(s) = p/q$, the island width is:
\begin{equation}
\Delta s = 4 \sqrt{\frac{\max |\nu|}{\omega_{\chi}'(s) 2\pi q }},
\label{eq:iw_passing}
\end{equation}
where $\max|\nu|$ refers to the maximum value of $|\nu|$ among all pseudo-orbits equally spaced in $\phi$ at the resonance $p/q$. For trapped particles with a resonance $\omega_{\zeta}(s) = p/q$, the island width is:
\begin{equation}
\Delta s = 4 \sqrt{\frac{\max |\nu|}{\omega_{\zeta}'(s) 2\pi q }}.
\label{eq:iw_trapped}
\end{equation}
The details of this derivation from the Hamiltonian in action-angle coordinates is provided in Appendix \ref{app:ham}, and application to guiding-center motion is shown in Appendix \ref{app:map}.

\begin{figure}
    \centering
    \includegraphics[width=0.8\textwidth]{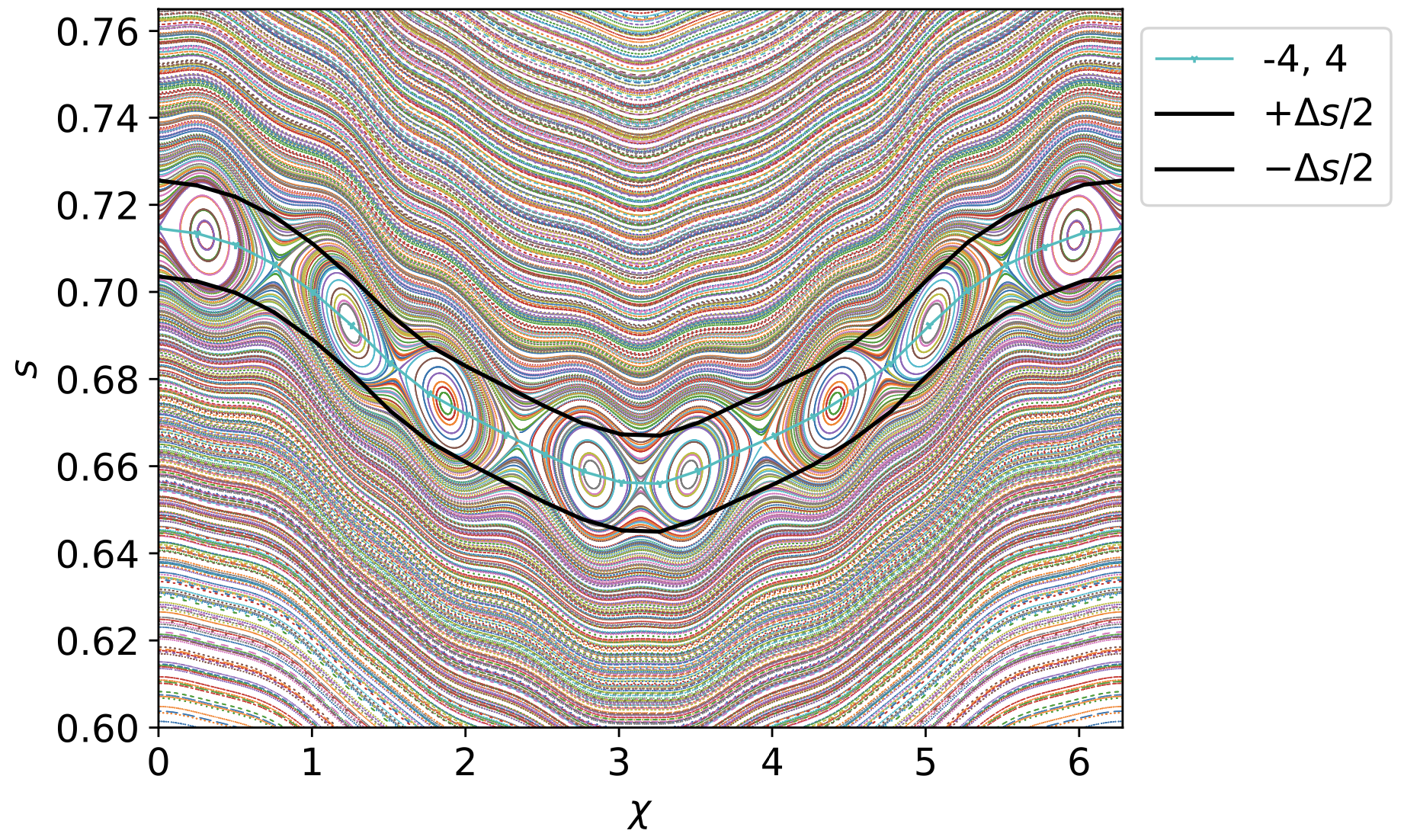}
    \caption{Comparison of theoretical island width with numerically determined result for co-propagating passing particles in QH. Theoretical island width is produced from (\ref{eq:iw_passing}) using the corresponding value of $\nu$ output by the pseudo-periodic curve routine.}
    \label{fig:iw_qhb_p}
\end{figure}

\section{Results}
\label{sec:results}
\begin{figure*}
    \centering
    \includegraphics[width=0.8\textwidth]{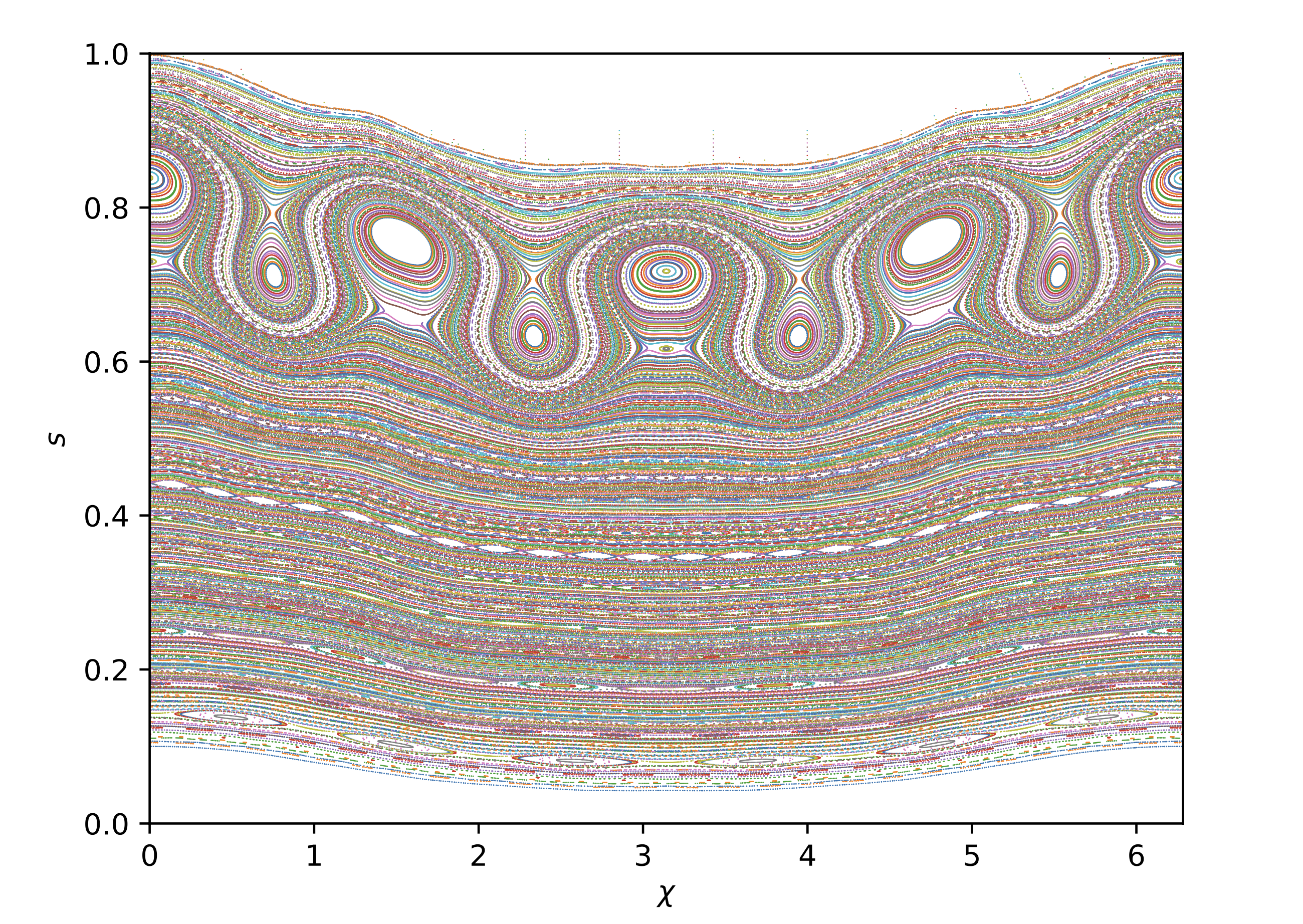}
    \caption{Poincar\'e map for co-propagating passing particles in QA $\beta=2.5\%$. The map demonstrates a structure characteristic of low shear in the region $0.6 < s < 0.9$.}
    \label{fig:qab_p}
\end{figure*}

\begin{figure*}
    \centering
    \begin{subfigure}[b]{0.49\textwidth}
        \centering
        \includegraphics[width=\textwidth]{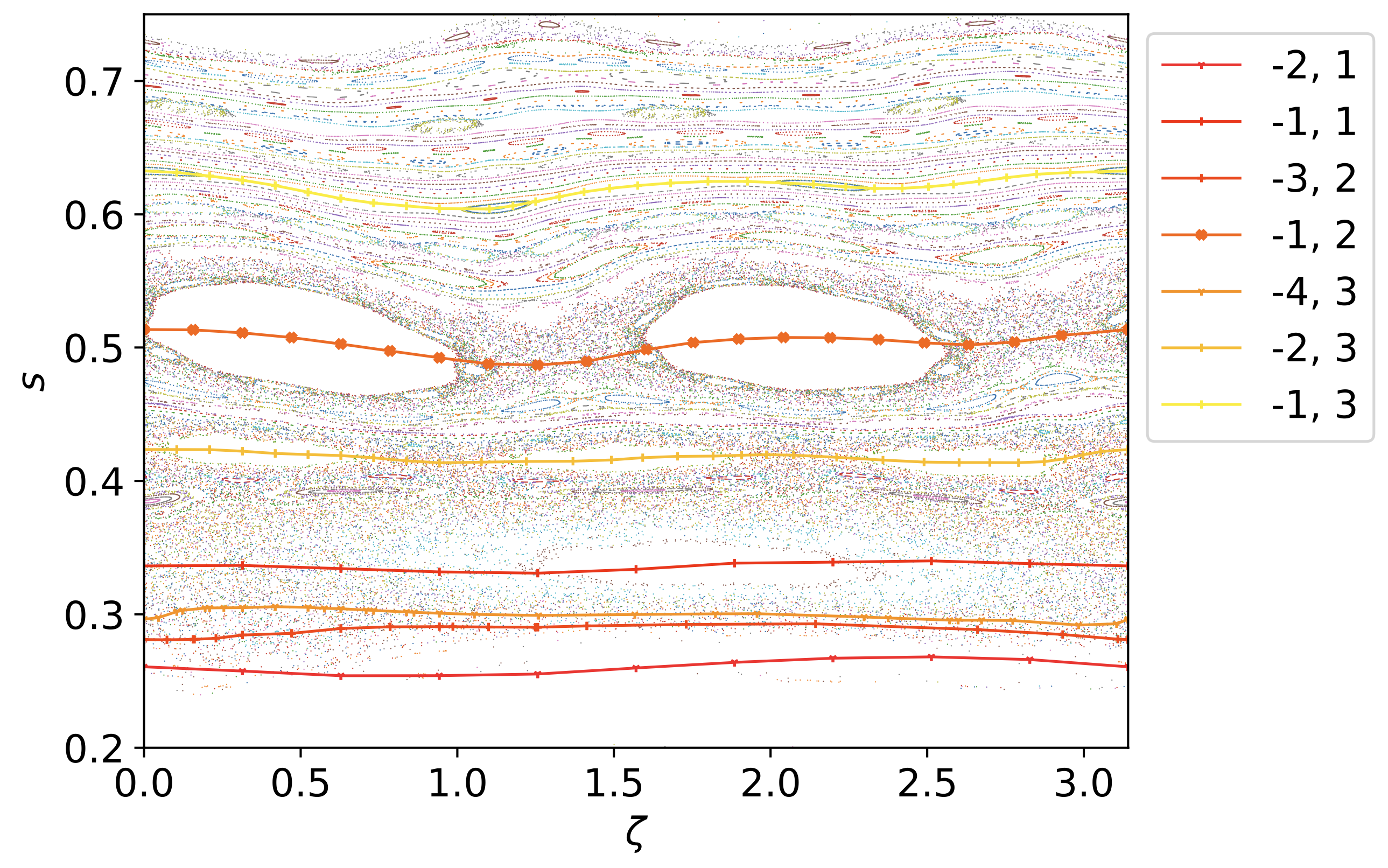}
        \caption{Poincar\'e map and set of pseudo-periodic curves for trapped particles in QA.}
        \label{fig:qab_t}
    \end{subfigure}
    \hfill
    \begin{subfigure}[b]{0.45\textwidth}
        \centering
        \includegraphics[width=\textwidth]{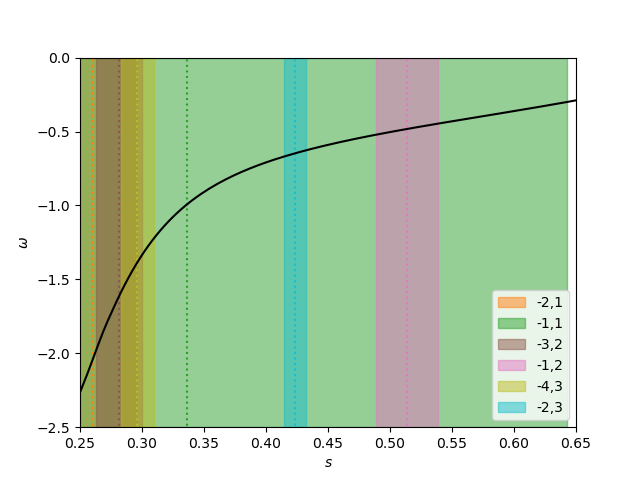}
        \caption{Island overlap and $\omega_{\zeta}$ for trapped particles in QA.}
        \label{fig:qab_t_iw}
    \end{subfigure}
    \caption{True map and set of pseudo-periodic curves for trapped particles in QA $\beta=2.5\%$ equilibrium, compared with island overlap for $\lambda=0.949$. Significant stochasticity in the true map is confirmed to be a result of island overlap, shown in \ref{fig:qab_t_iw} plotted with the toroidal transit frequency as a function of radius, shown in black. Island widths are computed from (\ref{eq:iw_trapped}). The radial location of each resonance is indicated by the dashed lines, while the width is given by the band in the corresponding color. The high degree of stochasticity is indicated by the many regions of island overlap across the domain.}
\end{figure*}

\indent We investigated characteristic frequencies and resonances crossings for fusion-born $\alpha$-particles in QA and QH equilibria. 
For trapped trajectories in the QH $\beta=2.5\%$ case shown in Figure \ref{fig:qhb_trapped}, multiple zero-crossings of the low shear toroidal precession frequency profile result in three distinct island structures. 
The resulting nontwist map \citep{DELCASTILLONEGRETE19961} caused by the low shear and non-monotonic behavior in $\omega_{\zeta}$ is shown in Figure \ref{fig:qhb_tr_freq}. Multiple island chains appear at different radial locations sharing the same X-points. These perturbations demonstrate the significant impact of the zero-crossing on wide island formation, supported by the analytic result presented in (\ref{eq:omega_zeta}). The shearless curve characteristic of nontwist maps is robust to perturbations \citep{mugnaine2024shearlesseffectivebarrierschaotic}, preventing stochasticity even for this wide island case. Equation (\ref{eq:omega_zeta}) shows that QH toroidal precession frequencies will be more sensitive to the zero-crossing than QA frequencies due to the inverse scaling with the square of effective helicity.\\
\indent For passing particles, agreement with (\ref{eq:iw_passing}) is investigated using the QH $\beta=2.5\%$ map for a $p=-1, q=2$ island chain shown in Figure \ref{fig:iw_qhb_p}. A set of pseudo-periodic curves are produced for the $(p,q)=(-1,2)$ resonance, shown in Figure \ref{fig:iw_qhb_p}. Given $\nu$ from the root-solve, we observe favorable agreement between the resulting island width and (\ref{eq:iw_passing}). This island chain is the only sizeable island structure for this case. The pseudo-periodic curve defined in Algorithm 1 successfully resolves the nearby integrable Hamiltonian system, passing through the X- and O-points of phase-space islands in the true map as expected. This demonstrates both the success of our pseudo-periodic curve construction and applicability of our analytic form for island width to characterize island formation.\\
\indent The Poincar\'e map for passing particles in QA is shown in Figure \ref{fig:qab_p}. A distinctive nontwist map structure is again observed. The topology of this case is indicated by the main island chain oscillating in radial position, as the X- and O-points are not collinear. Frequency behavior for this case is shown in Figure \ref{fig:freq_qab_p}, with low shear in the radial range $0.6<s<0.8$. The pseudo-periodic curve method is not robust to cases with large regions of low shear at a resonance, since degeneracy of resonance locations inhibits the radial root-solve.\\
\indent In contrast, the map for the trapped case in this QA equilibrium is highly stochastic, shown in Figure \ref{fig:qab_t}. The pseudo-periodic curves found indicate the presence of a number of low-order resonant perturbations in this configuration, implying susceptibility to large island formation. A high level of stochasticity is observed over the full domain, indicating diffusive transport for trapped particles in this configuration. The pseudo-periodic curves are used to quantify island width for each of these low-order resonances, and these widths are used to assess island overlap for this case. In Figure \ref{fig:qab_t_iw} we compare the phase-space map with (\ref{eq:iw_trapped}) to determine regions of island overlap for each low-order resonance. This extensive overlap results from this QA equilibrium having higher shear, as $\omega_{\zeta}$ crosses through several lower order resonances over a short radial distance and causing phase-space stochasticity. This suggests that minimizing the number of resonances crossed by energetic particle frequency profiles should be conducted for QA configurations. Additionally, the contribution of the perturbation from QA to the bounce-averaged radial drift over a resonant trajectory should be suppressed.\\
\indent Comparison between trapped and passing cases for both QA and QH reveals the difference in impact of resonances for each particle class. The map for passing trajectories in QA in Figure \ref{fig:iw_qhb_p} shows one significant resonance, while in contrast the trapped map in Figure \ref{fig:qab_t} demonstrates significant resonant overlap and the presence of several low-order resonances. For QH, the co-propagating passing map demonstrates a single resonance crossing with a fairly small island chain, while the trapped particle map at $\lambda=0.95$ demonstrates several large islands due to zero-crossing sensitivity. These representative cases indicate that the passing trajectories are less impacted by resonances than trapped trajectories. This is likely a result of the fact that equilibria are optimized to have few resonances for $\iota$, which is similar to $\omega_{\chi}$, as shown in Figure \ref{fig:all_frequencies}.

\section{Conclusion}
\label{sec:conclusion}
With phase-space Poincar\'e maps and quadratic flux-minimization, we use distance from integrability to characterize both diffusive and convective transport for trapped and passing guiding-center trajectories in both quasiaxisymmetric (QA) and quasihelical (QH) equilibria with $\beta=2.5\%$. We have shown that even in cases with low quasisymmetry error and where the magnetic field is integrable, phase-space integrability is not guaranteed. Characterization of island widths and island overlap agree favorably with numerical guiding-center tracing results from SIMSOPT. We observe that the QH equilibrium had characteristic frequency profiles with lower shear than the QA equilibrium for both particle classes. Our expression for trapped particle toroidal precession frequency (\ref{eq:omega_zeta}) using the near-axis formulation supports this result due to an inverse scaling with the square of effective helicity. The heightened sensitivity of trapped trajectories in QH to the zero-crossing of the precession frequency was indicated. As a result of lower shear, QH precession frequency profiles crossed through fewer low-order resonances than QA profiles. Correspondingly, the QH trapped particle toroidal precession frequency profile exhibited multiple zero-crossings and the resulting formation of large drift islands. Since this profile has low shear near zero, the zero-crossing is a more significant factor in convective transport than for the QA case considered. For trapped particles in QA, while this case had no zero-crossing, higher shear of the frequency profile contributed to several low-order resonance crossings, leading to wide overlapping islands forming, causing diffusive transport. Since QA equilibria tend to have higher precession frequency shear than QH equilibria, we conclude that island overlap and diffusive transport are more likely in QA than in QH equilibria. Additionally, the co-propagating passing map for finite-$\beta$ QA illustrated a characteristic nontwist map, corroborated by the non-monotonic frequency profile. Overall, we found that passing particle trajectories are less subject to resonances than trapped particle trajectories. This is likely a result of the fact that optimization to minimize resonances in $\iota$ can help reduce resonances in passing particle frequencies, though drifts can have some impact.
\\
\indent These methods can be used to evaluate equilibria at reactor scale. Future work could extend the QFM to be robust to nontwist map topologies. Evaluations of distance from integrability, island width, and island overlap could be used to inform equilibrium optimization. In particular, optimization of the drift-averaged frequency profiles over a range of pitch angles can be performed to alter shear profiles favorably \citep{Mackenbach}. For example, for QA, optimization could be performed to ensure the toroidal transit frequency profiles have lower shear to reduce the number of crossings through lower-order resonances, which could prevent overlap. For QH, the range of pitch angles for which the zero-crossing is present can be reduced. For both symmetries, careful quasisymmetry error-reduction could be conducted for particles over a range of pitch angles that resonate with low-order modes, and island width can be targeted. Using $\Gamma_c$, suberbanana resonances with small drift precession can targeted, but this does not address island width directly \citep{Velasco_2021}. By avoiding resonances through optimization, we expect to be less sensitive to non-quasisymmetric perturbations. This can be verified using sensitivity analysis with the shape gradient \citep{Landreman_2018} and Hessian matrix methods \citep{Zhu_2019}.\\
\indent There are many possible extensions of these methods. The non-conservation of the second adiabatic invariant, $J_{\|}$, can be shown to be correlated to stochastic loss mechanisms \citep{Paul_2022,Albert_2022}. Non-conservation of $J_{\|}$ could be investigated for cases with island overlap \citep{Albert_Buchholz_Kasilov_Kernbichler_Rath_2023}. The QFM method could also be applied to configurations further from quasisymmetry like NCSX and ARIES-CS \citep{NCSX,ARIES}. The pseudo-periodic curve routine fails for these equilibria further from quasisymmetry, so additional work can adapt these methods for cases with more quasisymmetry error. In addition, these methods could be adapted to investigate the impact of resonances on quasi-isodynamic and quasi-poloidal configurations.

\section{Acknowledgements}
This material is based upon work supported by the U.S. Department of
Energy, Office of Science, Office of Advanced Scientific Computing Research, Department of
Energy Computational Science Graduate Fellowship under Award Number DE-SC0024386. This report was prepared as an account of work sponsored by an agency of the
United States Government. Neither the United States Government nor any agency thereof, nor
any of their employees, makes any warranty, express or implied, or assumes any legal liability
or responsibility for the accuracy, completeness, or usefulness of any information, apparatus,
product, or process disclosed, or represents that its use would not infringe privately owned
rights. Reference herein to any specific commercial product, process, or service by trade name,
trademark, manufacturer, or otherwise does not necessarily constitute or imply its
endorsement, recommendation, or favoring by the United States Government or any agency
thereof. The views and opinions of authors expressed herein do not necessarily state or reflect
those of the United States Government or any agency thereof.

\newpage

\bibliographystyle{jpp}
\bibliography{jpp-instuctions}

\providecommand{\noopsort}[1]{}\providecommand{\singleletter}[1]{#1}%
\begin{thebibliography}{41}
\expandafter\ifx\csname natexlab\endcsname\relax\def\natexlab#1{#1}\fi
\def\au#1{#1} \def\ed#1{#1} \def\yr#1{#1}\def\at#1{#1}\def\jt#1{\textit{#1}} \def\bt#1{#1}\def\bvol#1{\textbf{#1}} \def\vol#1{#1} \def\pg#1{#1} \def\publ#1{#1}\def\arxiv#1{#1}\def\org#1{#1}\def\st#1{\textit{#1}}

\bibitem[Albert {\em et~al.\/}(2022)Albert, Rath, Babin, Buchholz, Kasilov \& Kernbichler]{Albert_2022}
{\sc \au{Albert, C.G.}, \au{Rath, K.}, \au{Babin, R.}, \au{Buchholz, R.}, \au{Kasilov, S.V.} \& \au{Kernbichler, W.}} \yr{2022}  \at{Resonant transport of fusion alpha particles in quasisymmetric stellarators}.  \jt{Journal of Physics: Conference Series}  \bvol{2397}~(1),  \pg{012009}.

\bibitem[Albert {\em et~al.\/}(2023)Albert, Buchholz, Kasilov, Kernbichler \& Rath]{Albert_Buchholz_Kasilov_Kernbichler_Rath_2023}
{\sc \au{Albert, Christopher~G.}, \au{Buchholz, Rico}, \au{Kasilov, Sergei~V.}, \au{Kernbichler, Winfried} \& \au{Rath, Katharina}} \yr{2023}  \at{Alpha particle confinement metrics based on orbit classification in stellarators}.  \jt{Journal of Plasma Physics}  \bvol{89}~(3),  \pg{955890301}.

\bibitem[Albert {\em et~al.\/}(2016)Albert, Heyn, Kapper, Kasilov, Kernbichler \& Martitsch]{NEORT}
{\sc \au{Albert, Christopher~G.}, \au{Heyn, Martin~F.}, \au{Kapper, Gernot}, \au{Kasilov, Sergei~V.}, \au{Kernbichler, Winfried} \& \au{Martitsch, Andreas~F.}} \yr{2016}  \at{{Evaluation of toroidal torque by non-resonant magnetic perturbations in tokamaks for resonant transport regimes using a Hamiltonian approach}}.  \jt{Physics of Plasmas}  \bvol{23}~(8),  \pg{082515},  \arxiv{arXiv: https://pubs.aip.org/aip/pop/article-pdf/doi/10.1063/1.4961084/16032726/082515\_1\_online.pdf}.

\bibitem[Bader {\em et~al.\/}(2019)Bader, Drevlak, Anderson, Faber, Hegna, Likin, Schmitt \& Talmadge]{Bader_Drevlak_Anderson_Faber_Hegna_Likin_Schmitt_Talmadge_2019}
{\sc \au{Bader, Aaron}, \au{Drevlak, M.}, \au{Anderson, D.~T.}, \au{Faber, B.~J.}, \au{Hegna, C.~C.}, \au{Likin, K.~M.}, \au{Schmitt, J.~C.} \& \au{Talmadge, J.~N.}} \yr{2019}  \at{Stellarator equilibria with reactor relevant energetic particle losses}.  \jt{Journal of Plasma Physics}  \bvol{85}~(5),  \pg{905850508}.

\bibitem[Beidler {\em et~al.\/}(2001)Beidler, Kolesnichenko, Marchenko, Sidorenko \& Wobig]{Diffusive}
{\sc \au{Beidler, C.~D.}, \au{Kolesnichenko, Ya.~I.}, \au{Marchenko, V.~S.}, \au{Sidorenko, I.~N.} \& \au{Wobig, H.}} \yr{2001}  \at{{Stochastic diffusion of energetic ions in optimized stellarators}}.  \jt{Physics of Plasmas}  \bvol{8}~(6),  \pg{2731--2738},  \arxiv{arXiv: https://pubs.aip.org/aip/pop/article-pdf/8/6/2731/19323267/2731\_1\_online.pdf}.

\bibitem[del Castillo-Negrete {\em et~al.\/}(1996)del Castillo-Negrete, Greene \& Morrison]{DELCASTILLONEGRETE19961}
{\sc \au{del Castillo-Negrete, D.}, \au{Greene, J.M.} \& \au{Morrison, P.J.}} \yr{1996}  \at{Area preserving nontwist maps: periodic orbits and transition to chaos}.  \jt{Physica D: Nonlinear Phenomena}  \bvol{91}~(1),  \pg{1--23}.

\bibitem[Dewar {\em et~al.\/}(2012)Dewar, Hudson \& Gibson]{DEWAR20122062}
{\sc \au{Dewar, R.L.}, \au{Hudson, S.R.} \& \au{Gibson, A.M.}} \yr{2012}  \at{Action-gradient-minimizing pseudo-orbits and almost-invariant tori}.  \jt{Communications in Nonlinear Science and Numerical Simulation}  \bvol{17}~(5),  \pg{2062--2073}, special Issue: Mathematical Structure of Fluids and Plasmas.

\bibitem[Dewar {\em et~al.\/}(1994)Dewar, Hudson \& Price]{DEWAR1994}
{\sc \au{Dewar, R.L.}, \au{Hudson, S.R.} \& \au{Price, P.F.}} \yr{1994}  \at{Almost invariant manifolds for divergence-free fields}.  \jt{Physics Letters A}  \bvol{194}~(1),  \pg{49--56}.

\bibitem[Dewar \& Khorev(1995)]{DEWAR199566}
{\sc \au{Dewar, R.L.} \& \au{Khorev, A.B.}} \yr{1995}  \at{Rational quadratic-flux minimizing circles for area-preserving twist maps}.  \jt{Physica D: Nonlinear Phenomena}  \bvol{85}~(1),  \pg{66--78}.

\bibitem[Garabedian(1996)]{QA}
{\sc \au{Garabedian, P.~R.}} \yr{1996}  \at{{Stellarators with the magnetic symmetry of a tokamak}}.  \jt{Physics of Plasmas}  \bvol{3}~(7),  \pg{2483--2485},  \arxiv{arXiv: https://pubs.aip.org/aip/pop/article-pdf/3/7/2483/12414492/2483\_1\_online.pdf}.

\bibitem[Goldston {\em et~al.\/}(1981)Goldston, White \& Boozer]{Goldston}
{\sc \au{Goldston, R.~J.}, \au{White, R.~B.} \& \au{Boozer, A.~H.}} \yr{1981}  \at{Confinement of high-energy trapped particles in tokamaks}.  \jt{Phys. Rev. Lett.}  \bvol{47},  \pg{647--649}.

\bibitem[Goodman {\em et~al.\/}(2024)Goodman, Xanthopoulos, Plunk, Smith, Nührenberg, Beidler, Henneberg, Roberg-Clark, Drevlak \& Helander]{goodman2024quasiisodynamicstellaratorslowturbulence}
{\sc \au{Goodman, Alan~G.}, \au{Xanthopoulos, Pavlos}, \au{Plunk, Gabriel~G.}, \au{Smith, Håkan}, \au{Nührenberg, Carolin}, \au{Beidler, Craig~D.}, \au{Henneberg, Sophia~A.}, \au{Roberg-Clark, Gareth}, \au{Drevlak, Michael} \& \au{Helander, Per}} \yr{2024} Quasi-isodynamic stellarators with low turbulence as fusion reactor candidates,  \arxiv{arXiv: 2405.19860}.

\bibitem[Helander(2014)]{Helander_2014}
{\sc \au{Helander, Per}} \yr{2014}  \at{Theory of plasma confinement in non-axisymmetric magnetic fields}.  \jt{Reports on Progress in Physics}  \bvol{77}~(8),  \pg{087001}.

\bibitem[Hudson \& Dewar(1996)]{Dewar_1995}
{\sc \au{Hudson, S.} \& \au{Dewar, Robert}} \yr{1996}  \at{Almost-invariant surfaces for magnetic field-line flows}.  \jt{Journal of Plasma Physics}  \bvol{56},  \pg{361 -- 382}.

\bibitem[Hudson \& Dewar(1998)]{HUDSON1998246}
{\sc \au{Hudson, S.R.} \& \au{Dewar, R.L.}} \yr{1998}  \at{Construction of an integrable field close to any non-integrable toroidal magnetic field}.  \jt{Physics Letters A}  \bvol{247}~(3),  \pg{246--251}.

\bibitem[Hudson \& Dewar(1999)]{HudsonandDewar1999}
{\sc \au{Hudson, S.~R.} \& \au{Dewar, R.~L.}} \yr{1999}  \at{{Analysis of perturbed magnetic fields via construction of nearby integrable fields}}.  \jt{Physics of Plasmas}  \bvol{6}~(5),  \pg{1532--1538},  \arxiv{arXiv: https://pubs.aip.org/aip/pop/article-pdf/6/5/1532/19073319/1532\_1\_online.pdf}.

\bibitem[Isaev {\em et~al.\/}(1999)Isaev, Mikhailov, Monticello, Mynick, Subbotin, Ku \& Reiman]{NCSX}
{\sc \au{Isaev, M.~Yu.}, \au{Mikhailov, M.~I.}, \au{Monticello, D.~A.}, \au{Mynick, H.~E.}, \au{Subbotin, A.~A.}, \au{Ku, L.~P.} \& \au{Reiman, A.~H.}} \yr{1999}  \at{{The pseudo-symmetric optimization of the National Compact Stellarator Experiment}}.  \jt{Physics of Plasmas}  \bvol{6}~(8),  \pg{3174--3179},  \arxiv{arXiv: https://pubs.aip.org/aip/pop/article-pdf/6/8/3174/12265267/3174\_1\_online.pdf}.

\bibitem[Jorge {\em et~al.\/}(2020)Jorge, Sengupta \& Landreman]{Jorge_Sengupta_Landreman_2020}
{\sc \au{Jorge, R.}, \au{Sengupta, W.} \& \au{Landreman, M.}} \yr{2020}  \at{Near-axis expansion of stellarator equilibrium at arbitrary order in the distance to the axis}.  \jt{Journal of Plasma Physics}  \bvol{86}~(1),  \pg{905860106}.

\bibitem[Kim {\em et~al.\/}(2013)Kim, Park \& Boozer]{PhysRevLett.110.185004}
{\sc \au{Kim, Kimin}, \au{Park, Jong-Kyu} \& \au{Boozer, Allen~H.}} \yr{2013}  \at{Numerical verification of bounce-harmonic resonances in neoclassical toroidal viscosity for tokamaks}.  \jt{Phys. Rev. Lett.}  \bvol{110},  \pg{185004}.

\bibitem[Ku {\em et~al.\/}(2008)Ku, Garabedian, Lyon, Turnbull, Grossman, Mau, Zarnstorff \& Team]{ARIES}
{\sc \au{Ku, L.~P.}, \au{Garabedian, P.~R.}, \au{Lyon, J.}, \au{Turnbull, A.}, \au{Grossman, A.}, \au{Mau, T.~K.}, \au{Zarnstorff, M.} \& \au{Team, ARIES}} \yr{2008}  \at{Physics design for aries-cs}.  \jt{Fusion Science and Technology}  \bvol{54}~(3),  \pg{673--693},  \arxiv{arXiv: https://doi.org/10.13182/FST08-A1899}.

\bibitem[Landreman {\em et~al.\/}(2022)Landreman, Buller \& Drevlak]{QA_config}
{\sc \au{Landreman, M.}, \au{Buller, S.} \& \au{Drevlak, M.}} \yr{2022}  \at{{Optimization of quasi-symmetric stellarators with self-consistent bootstrap current and energetic particle confinement}}.  \jt{Physics of Plasmas}  \bvol{29}~(8),  \pg{082501},  \arxiv{arXiv: https://pubs.aip.org/aip/pop/article-pdf/doi/10.1063/5.0098166/19814696/082501\_1\_online.pdf}.

\bibitem[Landreman {\em et~al.\/}(2021)Landreman, Medasani, Wechsung, Giuliani, Jorge \& Zhu]{SIMSOPT}
{\sc \au{Landreman, Matt}, \au{Medasani, Bharat}, \au{Wechsung, Florian}, \au{Giuliani, Andrew}, \au{Jorge, Rogerio} \& \au{Zhu, Caoxiang}} \yr{2021}  \at{Simsopt: A flexible framework for stellarator optimization}.  \jt{Journal of Open Source Software}  \bvol{6}~(65),  \pg{3525}.

\bibitem[Landreman \& Paul(2018)]{Landreman_2018}
{\sc \au{Landreman, Matt} \& \au{Paul, Elizabeth}} \yr{2018}  \at{Computing local sensitivity and tolerances for stellarator physics properties using shape gradients}.  \jt{Nuclear Fusion}  \bvol{58}~(7),  \pg{076023}.

\bibitem[Landreman \& Paul(2022)]{QH_vac}
{\sc \au{Landreman, Matt} \& \au{Paul, Elizabeth}} \yr{2022}  \at{Magnetic fields with precise quasisymmetry for plasma confinement}.  \jt{Phys. Rev. Lett.}  \bvol{128},  \pg{035001}.

\bibitem[Landreman \& Sengupta(2018)]{Landreman_Sengupta_2018}
{\sc \au{Landreman, Matt} \& \au{Sengupta, Wrick}} \yr{2018}  \at{Direct construction of optimized stellarator shapes. part 1. theory in cylindrical coordinates}.  \jt{Journal of Plasma Physics}  \bvol{84}~(6),  \pg{905840616}.

\bibitem[LeViness {\em et~al.\/}(2022)LeViness, Schmitt, Lazerson, Bader, Faber, Hammond \& Gates]{LeViness_2023}
{\sc \au{LeViness, Alexandra}, \au{Schmitt, John~C.}, \au{Lazerson, Samuel~A.}, \au{Bader, Aaron}, \au{Faber, Benjamin~J.}, \au{Hammond, Kenneth~C.} \& \au{Gates, David~A.}} \yr{2022}  \at{Energetic particle optimization of quasi-axisymmetric stellarator equilibria}.  \jt{Nuclear Fusion}  \bvol{63}~(1),  \pg{016018}.

\bibitem[Lichtenberg \& Lieberman(1983)]{Lichtenberg_Lieberman}
{\sc \au{Lichtenberg, A.J.} \& \au{Lieberman, M.A.}} \yr{1983} {\em Regular and Chaotic Dynamics\/}, second edition edn.  \publ{Springer-Verlag New York Inc.}

\bibitem[Littlejohn(1983)]{Littlejohn_1983}
{\sc \au{Littlejohn, Robert~G.}} \yr{1983}  \at{Variational principles of guiding centre motion}.  \jt{Journal of Plasma Physics}  \bvol{29}~(1),  \pg{111–125}.

\bibitem[Mackenbach {\em et~al.\/}(2023)Mackenbach, Duff, Gerard, Proll, Helander \& Hegna]{Mackenbach}
{\sc \au{Mackenbach, R. J.~J.}, \au{Duff, J.~M.}, \au{Gerard, M.~J.}, \au{Proll, J. H.~E.}, \au{Helander, P.} \& \au{Hegna, C.~C.}} \yr{2023}  \at{{Bounce-averaged drifts: Equivalent definitions, numerical implementations, and example cases}}.  \jt{Physics of Plasmas}  \bvol{30}~(9),  \pg{093901},  \arxiv{arXiv: https://pubs.aip.org/aip/pop/article-pdf/doi/10.1063/5.0160282/18114741/093901\_1\_5.0160282.pdf}.

\bibitem[Mau {\em et~al.\/}(2008)Mau, Kaiser, Grossman, Raffray, Wang, Lyon, Maingi, Ku, Zarnstorff \& Team]{heatflux}
{\sc \au{Mau, T.~K.}, \au{Kaiser, T.~B.}, \au{Grossman, A.~A.}, \au{Raffray, A.~R.}, \au{Wang, X.~R.}, \au{Lyon, J.~F.}, \au{Maingi, R.}, \au{Ku, L.~P.}, \au{Zarnstorff, M.~C.} \& \au{Team, ARIES-CS}} \yr{2008}  \at{Divertor configuration and heat load studies for the aries-cs fusion power plant}.  \jt{Fusion Science and Technology}  \bvol{54}~(3),  \pg{771--786},  \arxiv{arXiv: https://doi.org/10.13182/FST08-27}.

\bibitem[Mugnaine {\em et~al.\/}(2024)Mugnaine, au2, Viana, Caldas \& Morrison]{mugnaine2024shearlesseffectivebarrierschaotic}
{\sc \au{Mugnaine, M.}, \au{au2, J. D. Szezech~Jr.}, \au{Viana, R.~L.}, \au{Caldas, I.~L.} \& \au{Morrison, P.~J.}} \yr{2024} Shearless effective barriers to chaotic transport induced by even twin islands in nontwist systems,  \arxiv{arXiv: 2406.19947}.

\bibitem[Mynick(1993)]{mynick}
{\sc \au{Mynick, H.~E.}} \yr{1993}  \at{{Transport of energetic ions by low‐n magnetic perturbations}}.  \jt{Physics of Fluids B: Plasma Physics}  \bvol{5}~(5),  \pg{1471--1481},  \arxiv{arXiv: https://pubs.aip.org/aip/pfb/article-pdf/5/5/1471/12440341/1471\_1\_online.pdf}.

\bibitem[Nemov {\em et~al.\/}(2008)Nemov, Kasilov, Kernbichler \& Leitold]{Nemov}
{\sc \au{Nemov, V.~V.}, \au{Kasilov, S.~V.}, \au{Kernbichler, W.} \& \au{Leitold, G.~O.}} \yr{2008}  \at{{Poloidal motion of trapped particle orbits in real-space coordinates}}.  \jt{Physics of Plasmas}  \bvol{15}~(5),  \pg{052501},  \arxiv{arXiv: https://pubs.aip.org/aip/pop/article-pdf/doi/10.1063/1.2912456/14080658/052501\_1\_online.pdf}.

\bibitem[Park {\em et~al.\/}(2013)Park, Jeon, Menard, Ko, Lee, Bae, Joung, You, Lee, Logan, Kim, Ko, Yoon, Hahn, Kim, Kim, Oh \& Kwak]{PhysRevLett.111.095002}
{\sc \au{Park, J.-K.}, \au{Jeon, Y.~M.}, \au{Menard, J.~E.}, \au{Ko, W.~H.}, \au{Lee, S.~G.}, \au{Bae, Y.~S.}, \au{Joung, M.}, \au{You, K.-I.}, \au{Lee, K.-D.}, \au{Logan, N.}, \au{Kim, K.}, \au{Ko, J.~S.}, \au{Yoon, S.~W.}, \au{Hahn, S.~H.}, \au{Kim, J.~H.}, \au{Kim, W.~C.}, \au{Oh, Y.-K.} \& \au{Kwak, J.-G.}} \yr{2013}  \at{Rotational resonance of nonaxisymmetric magnetic braking in the kstar tokamak}.  \jt{Phys. Rev. Lett.}  \bvol{111},  \pg{095002}.

\bibitem[Paul {\em et~al.\/}(2022)Paul, Bhattacharjee, Landreman, Alex, Velasco \& Nies]{Paul_2022}
{\sc \au{Paul, E.J.}, \au{Bhattacharjee, A.}, \au{Landreman, M.}, \au{Alex, D.}, \au{Velasco, J.L.} \& \au{Nies, R.}} \yr{2022}  \at{Energetic particle loss mechanisms in reactor-scale equilibria close to quasisymmetry}.  \jt{Nuclear Fusion}  \bvol{62}~(12),  \pg{126054}.

\bibitem[Rodríguez \& Mackenbach(2023)]{Rodríguez_Mackenbach_2023}
{\sc \au{Rodríguez, E.} \& \au{Mackenbach, R.J.J.}} \yr{2023}  \at{Trapped-particle precession and modes in quasisymmetric stellarators and tokamaks: a near-axis perspective}.  \jt{Journal of Plasma Physics}  \bvol{89}~(5),  \pg{905890521}.

\bibitem[Velasco {\em et~al.\/}(2021)Velasco, Calvo, Mulas, Sánchez, Parra, Cappa \& the W7-X~Team]{Velasco_2021}
{\sc \au{Velasco, J.L.}, \au{Calvo, I.}, \au{Mulas, S.}, \au{Sánchez, E.}, \au{Parra, F.I.}, \au{Cappa, A.} \& \au{the W7-X~Team}} \yr{2021}  \at{A model for the fast evaluation of prompt losses of energetic ions in stellarators}.  \jt{Nuclear Fusion}  \bvol{61}~(11),  \pg{116059}.

\bibitem[White {\em et~al.\/}(2022)White, Bierwage \& Ethier]{White_2022}
{\sc \au{White, Roscoe}, \au{Bierwage, Andreas} \& \au{Ethier, Stephane}} \yr{2022}  \at{Poor confinement in stellarators at high energy}.  \jt{Physics of Plasmas}  \bvol{29}~(5),  \pg{052511},  \arxiv{arXiv: https://doi.org/10.1063/5.0094458}.

\bibitem[White(2011)]{White_2011}
{\sc \au{White, R~B}} \yr{2011}  \at{Modification of particle distributions by magnetohydrodynamic instabilities ii}.  \jt{Plasma Physics and Controlled Fusion}  \bvol{53}~(8),  \pg{085018}.

\bibitem[White {\em et~al.\/}(2015)White, Gates \& Brennan]{White_2015}
{\sc \au{White, R.~B.}, \au{Gates, D.~A.} \& \au{Brennan, D.~P.}} \yr{2015}  \at{Thermal island destabilization and the greenwald limit}.  \jt{Physics of Plasmas}  \bvol{22}~(2),  \pg{022514},  \arxiv{arXiv: https://doi.org/10.1063/1.4913433}.

\bibitem[Zhu {\em et~al.\/}(2019)Zhu, Gates, Hudson, Liu, Xu, Shimizu \& Okamura]{Zhu_2019}
{\sc \au{Zhu, Caoxiang}, \au{Gates, David~A.}, \au{Hudson, Stuart~R.}, \au{Liu, Haifeng}, \au{Xu, Yuhong}, \au{Shimizu, Akihiro} \& \au{Okamura, Shoichi}} \yr{2019}  \at{Identification of important error fields in stellarators using the hessian matrix method}.  \jt{Nuclear Fusion}  \bvol{59}~(12),  \pg{126007}.

\end{thebibliography}
\newpage
\section{Appendixes}
\appendix
\section{Characteristic frequency derivations}
\label{app:dzeta}

\subsection{Near-axis model overview}
The near-axis expansion can be used to provide an analytic form for the magnetic field from which we can produce an expression for trapped particle frequencies. This can provide insight into the behavior of trapped particles and their sensitivity to resonances. In the near-axis expansion, the magnetic field strength to first order in the distance from the magnetic axis $r = \sqrt{2\psi/B_0}$ is given by \citep{Jorge_Sengupta_Landreman_2020}
\begin{align}
    B = B_0 \left(1 - r \bar\eta \cos(\chi)\right),
\end{align}\label{eq:bfield_app}
where $\chi = \theta - N \zeta$ is the angle such that $B(r,\chi)$, $N$ is an integer which denotes the symmetry helicity, and $B_0$ and $\bar\eta$ are constants representing field strength on-axis and field strength variation, respectively. The Boozer covariant components and rotational transform are approximated as
\begin{align}
    G(r) &= G_0 + \mathcal{O}(r^2) \\ 
    I(r) &= I_2 r^2 + \mathcal{O}(r^4) \\
    \iota(r) &= \iota_0 + \mathcal{O}(r^2)
\end{align}
such that the field can be written as $\bm{B}=I(r)\nabla\theta + G(r)\nabla\zeta$.
\subsection{Characteristic frequencies}
To gain further insight into the sensitivity of trapped particles to different resonances, it is useful to develop an analytic expression for trapped particle precession frequency $\omega_{\zeta}$. Between bounce points, we average the precession over lowest order motion along the field line in $\rho_*$, so $\alpha=\theta-\iota\zeta$ is fixed to be constant over a trajectory. The precession frequency can then be related to change in poloidal angle by
\begin{equation}
    \omega_{\zeta}=-\frac{1}{2\pi(\iota_0-N)} \Delta \theta.
    \label{eq:dzeta_from_dtheta}
\end{equation}
For drift velocity $\boldsymbol{v}_d$ and unit field vector $\hat{\boldsymbol{b}}=\boldsymbol{B}/B$, the change in poloidal angle during one toroidal transit is
\begin{equation}
    \Delta \theta = \oint \frac{dl}{v_{\|}}\dot{\theta} = \frac{1}{\iota_0-N}\oint \frac{d\chi}{\hat{\bm{b}} \cdot \nabla \chi v_{\|}} \, \left(v_{\|} \hat{\bm{b}} +  \bm{v}_d\right) \cdot \nabla \theta
    \label{eq:theta}
\end{equation}
for integration performed in $\chi$ over a full bounce period. We can express $v_{\|}$ in terms of $B$ using the pitch angle $\lambda$:
\begin{align*}
    v_{\|}&=\pm v_0 \sqrt{1-\frac{\lambda B}{B_0}}.
\end{align*}
Using the contravariant form $\boldsymbol{B}=\nabla\psi\times\nabla\theta-\iota(\psi)\nabla\psi\times\nabla\zeta$ in (\ref{eq:theta}) gives
\begin{align}
    \Delta \theta &=\frac{2G_0}{\iota_0-N}\int_{\chi_-}^{\chi_+} \frac{d\chi}{B_0}\frac{v_{d,\theta}}{v_{\|}}
    \label{eq:deltatheta}
\end{align}
for integration between bounce points $\chi_-$ and $\chi_+$, since the sign of $v_{\|}$ changes depending on bounce direction, and the $\hat{\bm{b}}\cdot \nabla \theta$ term integrates to zero. We can determine $v_{d,\theta}$, the poloidal component of drift velocity, in the following way:
\begin{align*}
    \textbf{v}_d\cdot \nabla \theta&= \frac{\left(v_0^2+v_{\|}^2\right)}{2\Omega B^2} \left(\bm{B}\times \nabla B\right)\cdot \nabla \theta = \frac{-\bar\eta\left(v_0^2+v_{\|}^2\right)}{2\Omega r }\cos{\chi}
    \label{eq:vdotgradtheta}
\end{align*}
for $\Omega=qB_0/m$, the gyrofrequency. Substituting (\ref{eq:vdotgradtheta}) into (\ref{eq:deltatheta}),

\begin{align}
    \Delta \theta &= \frac{-\bar\eta G_0 v_0}{\Omega B_0 r}\frac{1}{\iota -N}\left[\int_{\chi_-}^{\chi_+} d\chi \frac{\cos{\chi}}{\sqrt{1-\lambda+\lambda r \bar\eta}}+\int_{\chi_-}^{\chi_+} d\chi \cos{\chi}\sqrt{1-\lambda+\lambda r \bar\eta}\right].
\end{align}
Through re-expression in terms of $k$ from (\ref{eq:k}), it becomes apparent that the second term is higher order in $r$. By bounce-point symmetry, to first order we have
\begin{align}
    \Delta \theta &= \frac{2\bar\eta G_0 v_0}{\Omega B_0 r}\frac{1}{\sqrt{2\lambda  r \bar\eta}}\frac{1}{\iota_0 -N}\int_{0}^{\chi_+} d\chi \frac{2\sin^2{\frac{\chi}{2}}-1}{\sqrt{k^2-\sin^2{\frac{\chi}{2}}}}
\end{align}
for
\begin{equation}
    k^2=\frac{1-\lambda+\lambda r \bar\eta}{2\lambda r \bar\eta}.
\end{equation}
Using the variable substitution $\phi=\frac{\chi}{2}$, followed by letting $\sin\phi=k\sin x$ and integrating over $x$ gives us our final expression for poloidal displacement:
\begin{align}
    \Delta \theta &=\frac{4\bar\eta G_0 v_0}{\Omega B_0 r}\frac{1}{\sqrt{2\lambda  r \bar\eta}}\frac{1}{\iota_0 -N}\left[2E(k^2)-K(k^2)\right].
\end{align}
Here $K(k^2)$ and $E(k^2)$ are elliptic integrals of the first and second kind, respectively. Using (\ref{eq:dzeta_from_dtheta}), we then have toroidal displacement
\begin{equation}
    \omega_{\zeta} =-\frac{2\bar\eta G_0 v_0}{\pi\Omega B_0 r}\frac{1}{\sqrt{2\lambda  r \bar\eta}}\frac{1}{(\iota_0 -N)^2}\left[2E(k^2)-K(k^2)\right].
\end{equation}

\subsection{Island width in Hamiltonian systems}
\label{app:ham}
In order to obtain expressions for island width in general action angle coordinates for guiding-center motion, we will make use of Hamiltonian formalism. This can be achieved if we treat the mapping integration angle $\rho$ as a time-like coordinate. For trapped particles, $\Delta \rho = 2\pi$ is the change in bounce angle after a bounce period, while for passing particles $\Delta \rho=2\pi$ is the change in $\zeta$ after a toroidal transit. We can then define the linear increase in canonical angle $\overline{\phi}=\omega_{\phi} \rho$ for $\phi=\{\chi,\zeta\}$ in terms of characteristic frequencies $\omega_{\phi}=\{\omega_{\chi},\omega_{\zeta}\}$ from (\ref{eq:wchi}) and (\ref{eq:wzeta}) \citep{Lichtenberg_Lieberman}. This can be used to develop an expression for drift island width. Consider a general one-dimensional Hamiltonian in action-angle coordinates $(J,\overline{\phi})$, where the unperturbed Hamiltonian reads $H_0(J)$. The unperturbed frequency is defined by $\omega_{\phi} = \partial H_0(J)/\partial J$. The perturbed Hamiltonian is then expressed in the action-angle coordinates of the unperturbed Hamiltonian as \citep{Lichtenberg_Lieberman}
\begin{align}
    H(J,\overline{\phi}) = H_0(J) + \sum_{p,q} H_{p,q}(J) \cos(q\overline{\phi} - p \rho). 
\end{align}
A resonance will occur if the following condition is satisfied:
\begin{align}
    q \omega_{\phi}(J_0) - p = 0.
    \label{eq:res_cond_app}
\end{align}
Assume that $J_0$ satisfies such a resonance condition. We can consider the transformed coordinates, $\Phi = \overline{\phi} - \omega_{\phi}(J_0)t$, for which the equations of motion read:
\begin{align}
\dot{\Phi} &=  \omega_{\phi}(J) - \omega_{\phi}(J_0) \\
\dot{J} &= \sum_{p,q} H_{p,q}(J) q  \sin\left( q  \left(\Phi + \omega_{\phi}(J_0) \rho \right)\right). 
\end{align}
For such a resonance, the island width in the $J$ action can then be evaluated as \citep{Lichtenberg_Lieberman}
\begin{align}
    \Delta J = 4\sqrt{\frac{H_{p,q}(J_0)}{\partial \omega_{\phi}(J_0)/\partial J}}.
    \label{eq:delta_J}
\end{align}

In the computation of $\omega_{\phi}$, trajectories are initialized at the intersection of $J$ with constant $\phi$, enabling a one-to-one mapping from $(J,\overline{\phi}) \rightarrow (s, \phi)$.

\section{Poincare map from Hamiltonian}
\label{app:map}
We can adapt the island width calculation in action-angle coordinates provided in Appendix \ref{app:ham} to the Poincar\'e maps we develop for this problem for island width analysis. In order to visualize the resonances in our system, we can make use of Poincar\'e maps defined for the different particle classes. In action-angle coordinates for the $n^{th}$ mapping, the Poincar\'{e} map is defined as $M(J^n,\overline{\phi}^n) \rightarrow (J^{n+1},\overline{\phi}^{n+1})$. In the integrable case,
\begin{align}
    J^{n+1} &= J^n \\
    \overline{\phi}^{n+1} &= \overline{\phi}^n + \Omega_{\rho}(J^{n+1}),
\end{align}
where action-angle displacement $\Omega_{\rho}(J^{n+1}) = \omega_{\phi}(J^{n+1}) \Delta \rho$. Here, $\omega_{\phi}(J^{n+1})$ is the unperturbed frequency and $\Delta \rho$ is the displacement in $\rho$ between successive mappings for integration angle $\rho$. For the resonance condition (\ref{eq:res_cond_app}), the orbit will close after $q$ applications of the map. In the near-integrable case, the twist mapping will be perturbed:
\begin{align}
    J^{n+q} &= J^n + f(J^{n+q},\overline{\phi}^n) \\
    \overline{\phi}^{n+q} &= \overline{\phi}^n + q\Omega_{\rho}(J^{n+q}) + g(J^{n+q},\overline{\phi}^n),
\end{align}
for some periodic functions $f$ and $g$. In the pseudo-periodic curve construction, we seek orbits that close in angle, so $g = 0$ and $f$ only depends on $\overline{\phi}$ and can be computed from the linearized Hamiltonian. After $q$ map applications,
\begin{align*}
    f &=\int_{\rho^n}^{\rho^{n+1}} d\rho \, \frac{dH_{p,q}(\boldsymbol{J_0})}{d\rho} = qH_{p,q}(\boldsymbol{J_0}) \sin\left(q\overline{\phi}^n-p\rho^n\right)\Delta \rho. 
\end{align*}
The distance between the periodic curve and the pseudo-periodic curve is quantified by the parameter $\nu = - f(J^{n+q},\overline{\phi}^n)$, as discussed in Section \ref{sec:qfm}.
We can then use $\nu$ to quantify island width in the following way:
\begin{align*}
\Delta J = 4 \sqrt{\frac{\max |\nu|}{\omega_{\phi}'(J) 2\pi q }},
\end{align*}
where max $|\nu|$ is the maximum value of $f$ taken over all trajectories in $\overline{\phi}$.
Since we initialize all particles at a constant angle $\overline{\phi}$, $J$ can be approximated as the flux function $s$. For the passing map analysis, $\omega_{\phi} \rightarrow \omega_{\chi}$.
\end{document}